\documentclass{article}
\PassOptionsToPackage{numbers,compress}{natbib}
\usepackage{preprint_style}
\usepackage[utf8]{inputenc}
\usepackage[T1]{fontenc}
\usepackage{hyperref}
\usepackage{url}
\usepackage{booktabs}
\usepackage{amsfonts}
\usepackage{amsmath}
\usepackage{amssymb}

\usepackage{nicefrac}
\usepackage{microtype}
\usepackage{xcolor}
\usepackage{graphicx}
\usepackage{subcaption}
\usepackage{multirow}
\usepackage{array}
\usepackage{siunitx}
\usepackage{enumitem}
\usepackage{float}
\usepackage{subcaption}
\usepackage{float}
\usepackage{soul}
\usepackage{textcomp}

\title{Physics-Grounded Understanding of Thermal Boundary Conductance between Ga$_2$O$_3$ and SiC from a Feedforward Neural Network Potential}
\author{
Nuohao Liu \\
Futurewei Technologies \\
University of Wisconsin--Madison \\
\texttt{nliu77@wisc.edu}
\And
Chen Shen \\
University of Wisconsin--Madison \\
\And
Yue Cao \\
Futurewei Technologies \\
\And
Song Xue \\
Futurewei Technologies \\
\AND
Pingfan Wu \\
Futurewei Technologies \\
\And
Zongfang Lin \\
Futurewei Technologies \\
\And
Masood Mortazavi\\
Futurewei Technologies \\
\And
Liang Peng\\
Futurewei Technologies \\
\AND
Izabela Szlufarska \\
University of Wisconsin--Madison \\
\And
Jiechen Wang \thanks{Corresponding author} \\
Futurewei Technologies \\
\texttt{jwang6@futurewei.com}
}

\begin{document}
\maketitle
\begin{abstract}
Ga$_2$O$_3$/SiC heterointegration is attractive for ultra-wide-bandgap power electronics, but interfacial thermal boundary conductance (TBC) remains a major heat-removal bottleneck. Direct experimental access to intrinsic atomistic interfacial transport remains limited, particularly for ideally synthesized materials with defect-free interfacial contact. First-principles simulations are too expensive at relevant length and time scales, while empirical Molecular Dynamics (MD) potentials often lack transferability across oxide and carbide bonding environments. We develop a unified feedforward neural network potential and validate it against density-functional data, bulk phonon dispersions, and anisotropic thermal-conductivity trends in both $\beta$-Ga$_2$O$_3$ and SiC. Nonequilibrium simulations show that TBC decreases with transport length, increases with temperature, and is consistently higher for Ga$_2$O$_3$$(\bar{2}01)$/SiC(0001) than for Ga$_2$O$_3$(100)/SiC(0001). These trends are explained by attenuation of long-mean-free-path carriers, enhanced incoherent and anharmonic interfacial exchange within broadly unchanged spectral channels, and stronger bonding and vibrational coupling at the $(\bar{2}01)$ interface. The results show how a single transferable feedforward neural network potential can enable large-scale transport prediction and physics-grounded mechanistic understanding of thermal boundary conductance. Code for NEP training and simulation workflows is available at the project repository \url{https://github.com/knowhow07/TBC_Ga2O3_SiC.git}
\end{abstract}

\section{Introduction}

\(\beta\)-Ga\(_2\)O\(_3\) is one of the most promising ultrawide-bandgap semiconductors for next-generation power and radio-frequency electronics because of its wide bandgap (\(\sim\)4.8--4.9 eV), high theoretical critical breakdown field (approaching 8 MV/cm), and scalability through melt-based bulk crystal growth \citep{Pearton2018-pj, Green2022-ah, Pearton2018-nf, Yuan2021-vv}. Its practical performance, however, is strongly limited by intrinsically low and anisotropic thermal conductivity (\(\sim 10\)--27 W\,m\(^{-1}\)\,K\(^{-1}\)), which causes severe self-heating under high-power operation \citep{Guo2015-Ga}. Heterogeneous integration with high-thermal-conductivity substrates is therefore a central thermal-management strategy, and 4H-SiC is especially attractive because of its high thermal conductivity, thermal stability, and relatively favorable lattice compatibility with Ga\(_2\)O\(_3\) \citep{Pearton2018-pj, Yuan2021-vv, Nepal2020-TBC-exp, Wei2013-rs, Qian2017-SiC}.

Despite this motivation, thermal transport across the Ga\(_2\)O\(_3\)/SiC interface remains incompletely understood. Experiments show that interfacial thermal boundary conductance (TBC) is highly sensitive to bonding and interfacial chemistry: MBE-grown \(\beta\)-Ga\(_2\)O\(_3\) on 4H-SiC exhibited a TBC of about \(140 \pm 60\) MW\,m\(^{-2}\)\,K\(^{-1}\) with a thin interfacial SiO\(_x\) layer, whereas a covalently bonded Ga\(_2\)O\(_3\)/SiO\(_2\)/SiC interface reached 162 MW\,m\(^{-2}\)\,K\(^{-1}\) \citep{Nepal2020-TBC-exp, Shen2025-NatCom}. First-principles studies likewise showed strong termination dependence, with Si--O configurations giving favorable stability and bonding, while recent molecular-dynamics studies examined heat transfer across related bonded and direct-contact heterointerfaces \citep{Xu2023-bond, Xu2023-interface, Shen2025-NatCom, Zhang2026-IJHMT}. Together, these results identify interfacial bonding, termination, and vibrational coupling as key factors, but a unified and physically transparent understanding of direct Ga\(_2\)O\(_3\)/SiC TBC, especially its dependence on transport length, temperature, and interface orientation, is still lacking. This makes Ga$_2$O$_3$/SiC interfacial thermal conductance a representative AI-for-materials problem: the relevant physics depends simultaneously on chemistry, lattice dynamics, and mesoscale transport length, which are prohibitively expensive to resolve with DFT alone and difficult to capture reliably with conventional empirical potentials.

Here we address this gap with a machine-learning approach aimed not only at computational efficiency but also at mechanism-resolving simulation across bulk-to-interface transport scales. We develop a compact yet transferable Ga--O--Si--C neuroevolution potential (NEP) that covers bulk, distorted, and interfacial environments within a single model. Using this validated potential, we perform systematic nonequilibrium simulations and mechanistic analysis to explain the length-, temperature-, and orientation-dependent TBC. Our main contribution is therefore a transport-faithful machine-learned interatomic potential for a chemically heterogeneous oxide/carbide interface, together with a physics-grounded analysis showing how carrier supply, anharmonic interfacial exchange, and orientation-dependent bonding jointly govern TBC in Ga$_2$O$_3$/SiC.

\section{Neural network potential training and simulation methodology}

\subsection{Neuroevolution Potential development and validation workflow}

For interfacial heat transport, first-principles calculations are prohibitively expensive at the required length and time scales, while empirical molecular-dynamics potentials often lack transferability across the distinct oxide and carbide bonding environments in Ga$_2$O$_3$/SiC. We therefore adopt a machine-learned interatomic potential. In particular, the neuroevolution potential (NEP) is well suited to thermal-transport studies because its compact single-hidden-layer feedforward architecture offers a favorable balance of accuracy and efficiency and its suitability for Ga$_2$O$_3$ thermal properties is further supported by prior NEP-based studies \citep{Fan2021-NEP,Fan2022,Sun2023-Ga$_2$O$_3$}.

\begin{figure}[t]
    \centering

    \includegraphics[width=\linewidth]{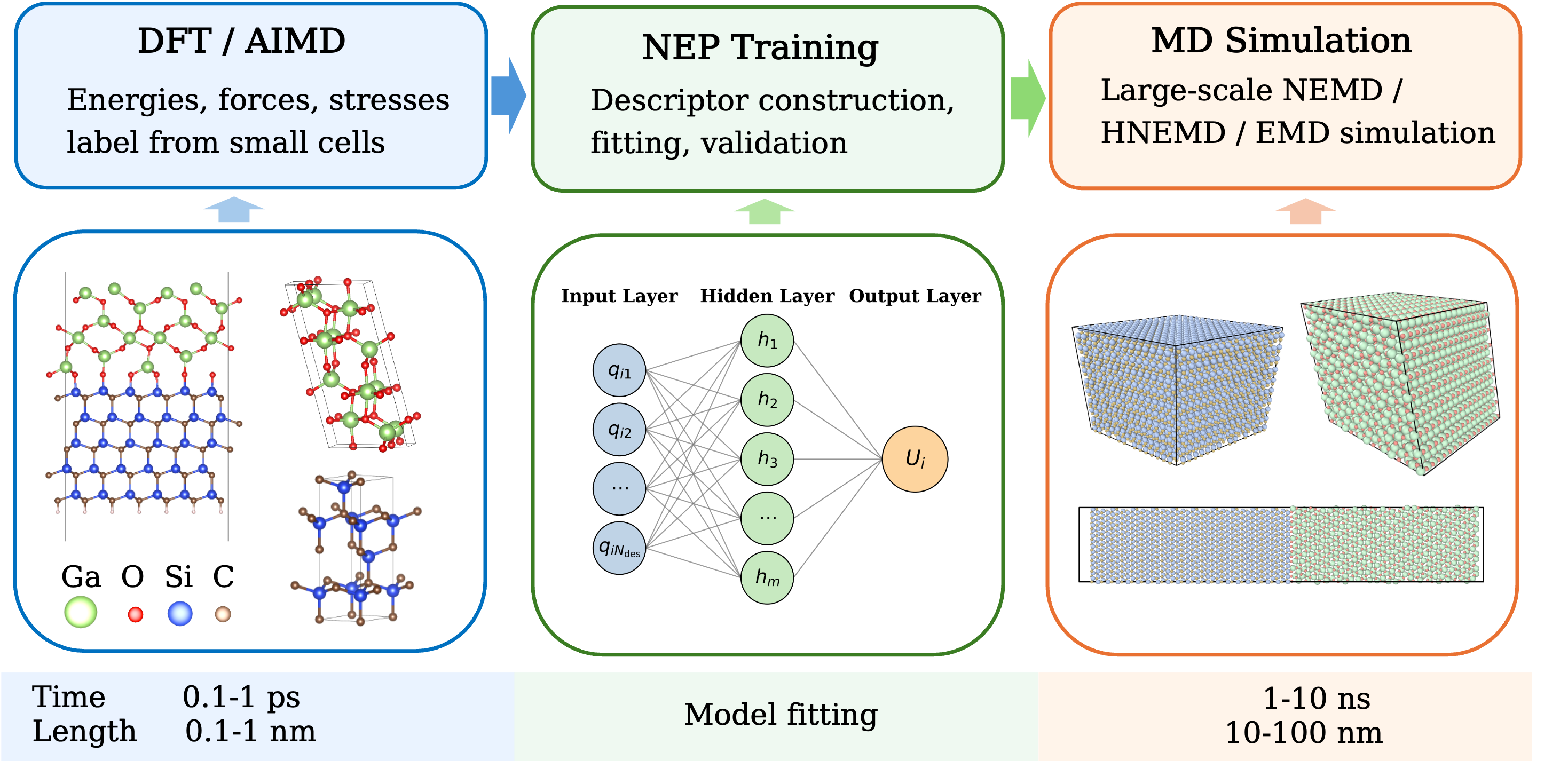}
    \caption{Overview of the multiscale workflow used in this work. DFT and AIMD calculations first provide reference energies, forces, and stress labels, which are used to construct descriptors and train the NEP model. The trained NEP is then applied in large-scale MD simulations, including NEMD, HNEMD, and EMD, to evaluate interfacial thermal transport. The bottom labels summarize the associated increase in characteristic time and length scales.}
    \label{fig1}
\end{figure}

Figure~\ref{fig1} summarizes the overall workflow. DFT and \textit{ab initio} molecular dynamics (AIMD) first provide reference energies, forces, and stresses from small representative cells, which are then used to train a unified Ga--O--Si--C NEP based on a compact feedforward neural-network architecture that maps local atomic descriptors to atomic energies. The trained potential is subsequently employed in large-scale molecular dynamics simulations, including equilibrium MD (EMD), nonequilibrium MD (NEMD), and homogeneous nonequilibrium MD (HNEMD), to reach the length and time scales required for bulk and interfacial thermal-transport analysis. A full list of abbreviations and explanations is provided in Appendix~\ref{app:abbrev}.

The training dataset was constructed to span the structural, chemical, and vibrational environments relevant to both bulk and interfacial transport, including equilibrium, strained and thermally perturbed snapshots for 4H-SiC, $\beta$-Ga$_2$O$_3$ and interface-containing structures. In total, about 30k DFT-labeled configurations were generated and split into training, validation, and test sets so that both near-equilibrium bulk vibrational environments and off-equilibrium local configurations near the interface are represented. A unified NEP was then trained on this dataset using a five-species type map \{Ga, O, Si, C, H\}, where H atoms are included only to passivate dangling bonds on the bottom free surface and stabilize the slab configurations \citep{Xu2023-bond}. In the NEP framework, the local environment of atom \(i\) is encoded by a descriptor vector \(\mathbf{q}_i=[q_{i1},q_{i2},\dots,q_{iN_{\mathrm{des}}}]\) containing radial and angular terms within prescribed cutoffs, and this descriptor is passed through a compact single-hidden-layer feedforward neural network in GPUMD v5.0 \citep{Fan2022} to predict the atomic site energy,
\begin{equation}
U_i = U_i(\mathbf{q}_i),
\end{equation}
which, for a hidden layer with \(N_{\mathrm{neu}}\) neurons, is written as
\begin{equation}
U_i =
\sum_{\mu=1}^{N_{\mathrm{neu}}}
w^{(1)}_{\mu}
\tanh\!\left(
\sum_{\nu=1}^{N_{\mathrm{des}}}
w^{(0)}_{\mu\nu} q^i_{\nu}
-
b^{(0)}_{\mu}
\right)
-
b^{(1)}.
\end{equation}
The total potential energy is then
\begin{equation}
E_{\mathrm{tot}} = \sum_i U_i .
\end{equation}
This shallow architecture combines physically informed descriptors with a compact nonlinear mapping, enabling a single potential to represent equilibrium bulk structures, thermally distorted configurations, and heterogeneous interface environments. Model parameters are optimized by minimizing a weighted multi-objective loss, which in simplified form is $\mathcal{L}=\lambda_E\mathrm{RMSE}(E)+\lambda_F\mathrm{RMSE}(F)+\lambda_V\mathrm{RMSE}(V)+\mathrm{reg.}$, where the terms correspond to root mean square errors (RMSE) in energies, forces, and virials, with optional $L_1$ and $L_2$ regularization \citep{Schaul2011}. The explicit loss function and full hyperparameters are provided in Appendix~\ref{app:nep-loss} and Appendix~\ref{app:nep-training}. Given the chemical species and atomic positions, the trained NEP predicts the total energy as a sum of local atomic contributions, with forces and stresses obtained from analytical energy derivatives. 

We validate the model at two levels. First, standard supervised accuracy is assessed using training curves and parity plots for energies and forces. Second, and more importantly, physics-based validation is performed by comparing bulk phonon dispersions and thermal conductivities of SiC and Ga$_2$O$_3$. This second level is essential because good parity statistics alone do not guarantee correct vibrational or transport behavior.

\subsection{MD simulation and transport mechanistic analysis} \label{subsec:NEMD}

We construct relaxed Ga$_2$O$_3$/SiC interface models for the $(100)$/SiC(0001) and $(\bar{2}01)$/SiC(0001) orientations and impose hot and cold regions at opposite ends of the simulation bar to establish a steady heat flux. These two orientations were selected based on prior studies identifying them as representative stable Ga$_2$O$_3$/SiC interface configurations \citep{Xu2023-bond,Xu2023-interface}. The thermal boundary conductance is computed as \citep{El_Hajj2024-es}
\begin{equation}
G=\frac{J}{\Delta T_{\mathrm{int}}},
\end{equation}
where \(J\) is the steady-state heat flux and \(\Delta T_{\mathrm{int}}\) is the temperature drop across the interface extracted from the NEMD temperature profile. Simulation implementation details are given in Appendix~\ref{app:sim}.

To analyze the microscopic origin of the conductance trends, we evaluate frequency-resolved spectral conductance, cross-interface coherence, and interfacial force-constant metrics. Cross-interface coherence is quantified by the magnitude-squared coherence function, and static interfacial coupling is characterized by the norm of the cross-interface harmonic interatomic force constant (IFC) blocks \citep{Carter1977-bs, Latour2014-lm}. These quantities probe complementary aspects of interfacial heat transfer, namely spectral transmission, vibrational phase correlation, and harmonic coupling strength. Full definitions are given in Appendix~\ref{app:equations}.

\begin{figure}[!t]
    \centering
    \begin{subfigure}[t]{0.37\linewidth}
        \centering
        \includegraphics[width=\linewidth]{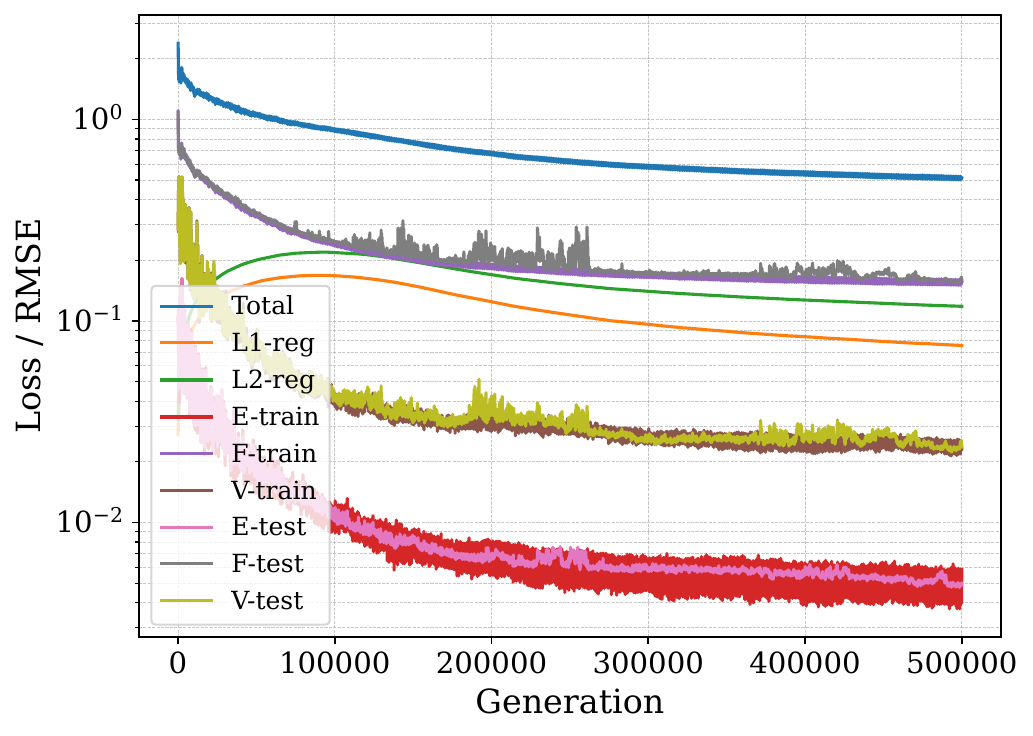}
        \caption{}
        \label{fig2:loss}
    \end{subfigure}
    \begin{subfigure}[t]{0.29\linewidth}
        \centering
        \includegraphics[width=\linewidth]{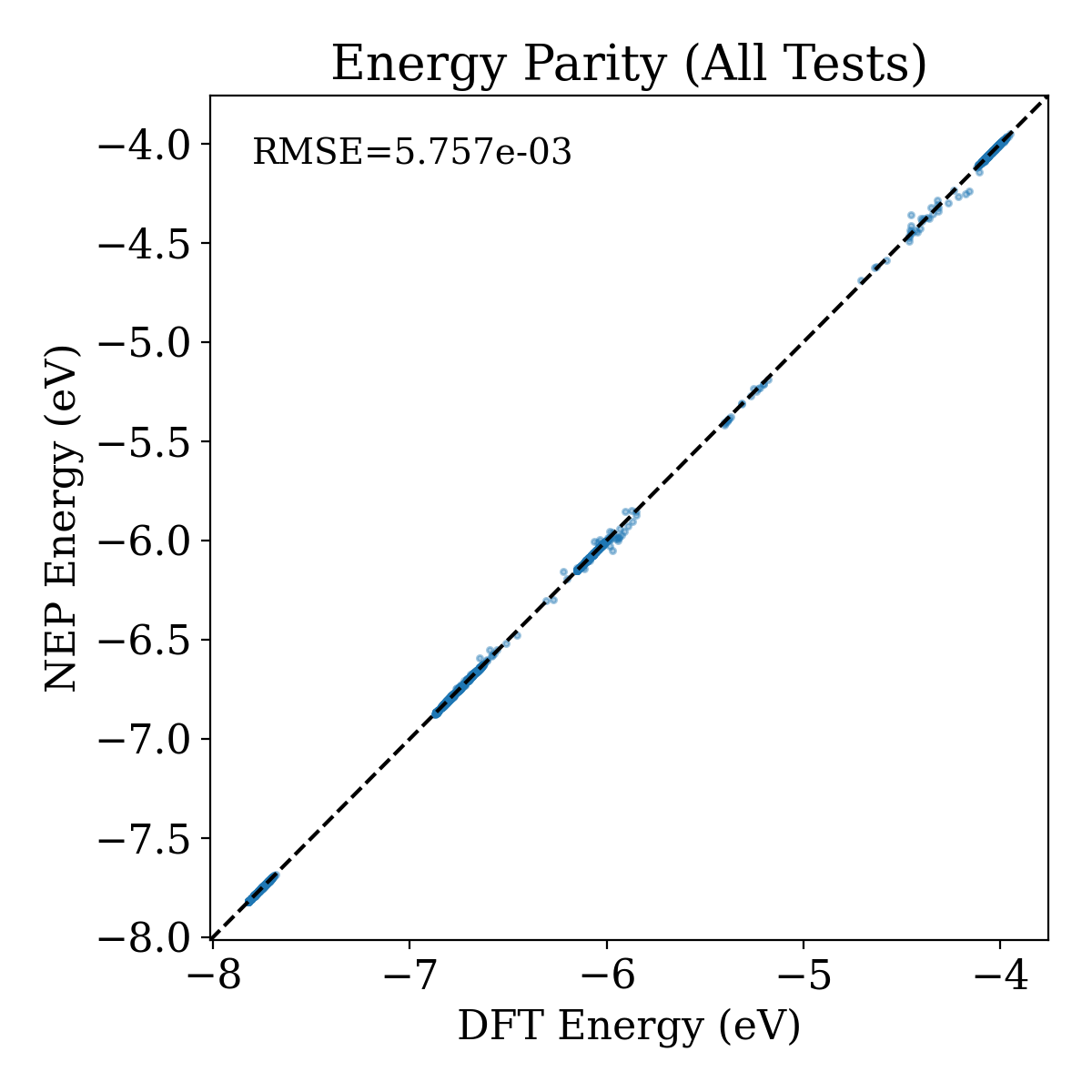}
        \caption{}
        \label{fig:energy}
    \end{subfigure}
    \begin{subfigure}[t]{0.29\linewidth}
        \centering
        \includegraphics[width=\linewidth]{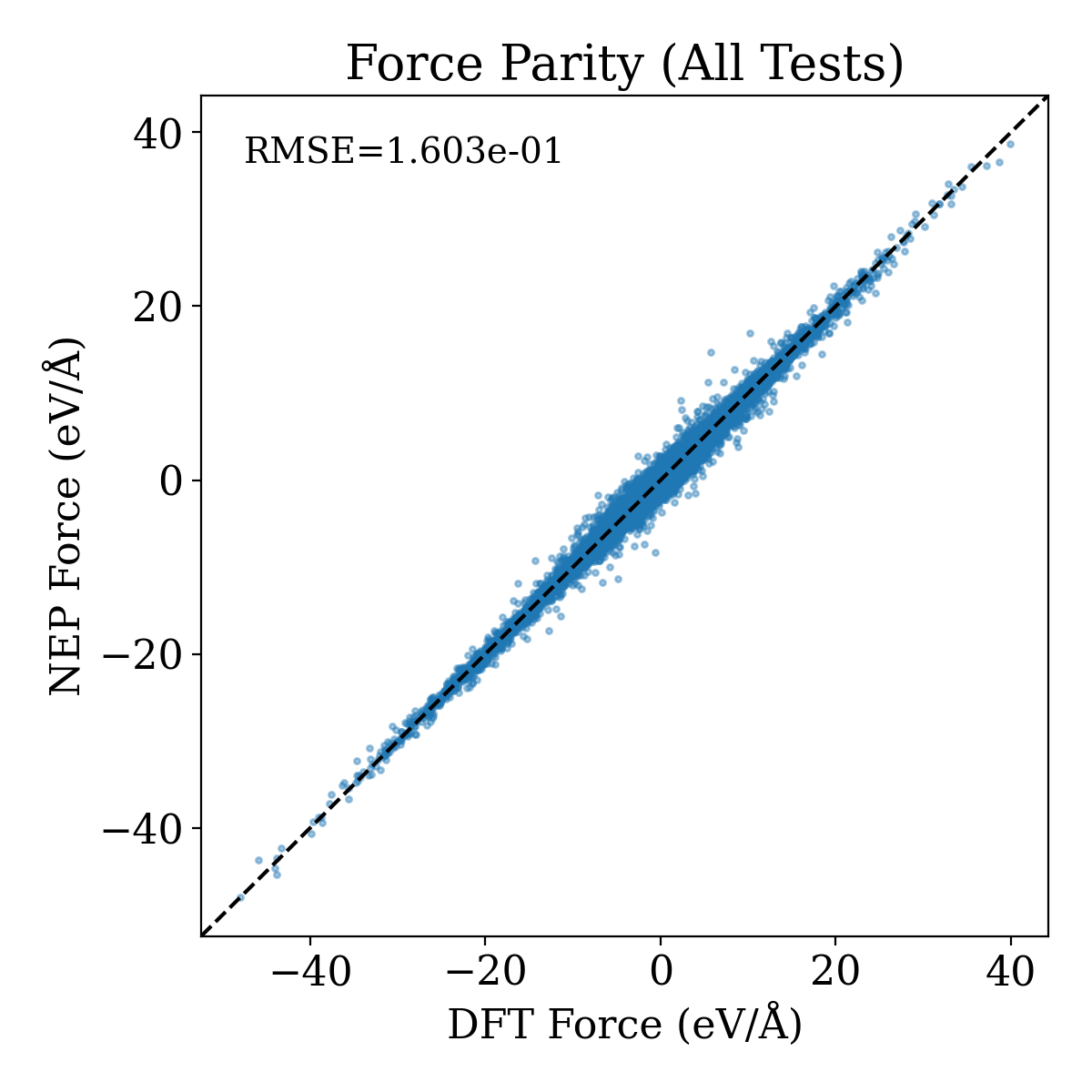}
        \caption{}
        \label{fig:force}
    \end{subfigure}
    \caption{ Supervised validation of the unified NEP. (a) Training loss curve. (b) Parity plot for energies. (c) Parity plot for forces. The smooth convergence and near-diagonal parity plots indicate that a single NEP can accurately represent the bulk and interfacial environments relevant to this work.}
    \label{fig2:nep-results}
\end{figure}

\section{Results and Physical Interpretation}
\subsection{Validation for NEP from fitting to transport-relevant bulk physics}

\begin{figure}[!htbp]
    \centering
    \begin{subfigure}[t]{0.45\linewidth}
        \centering
        \includegraphics[width=\linewidth]{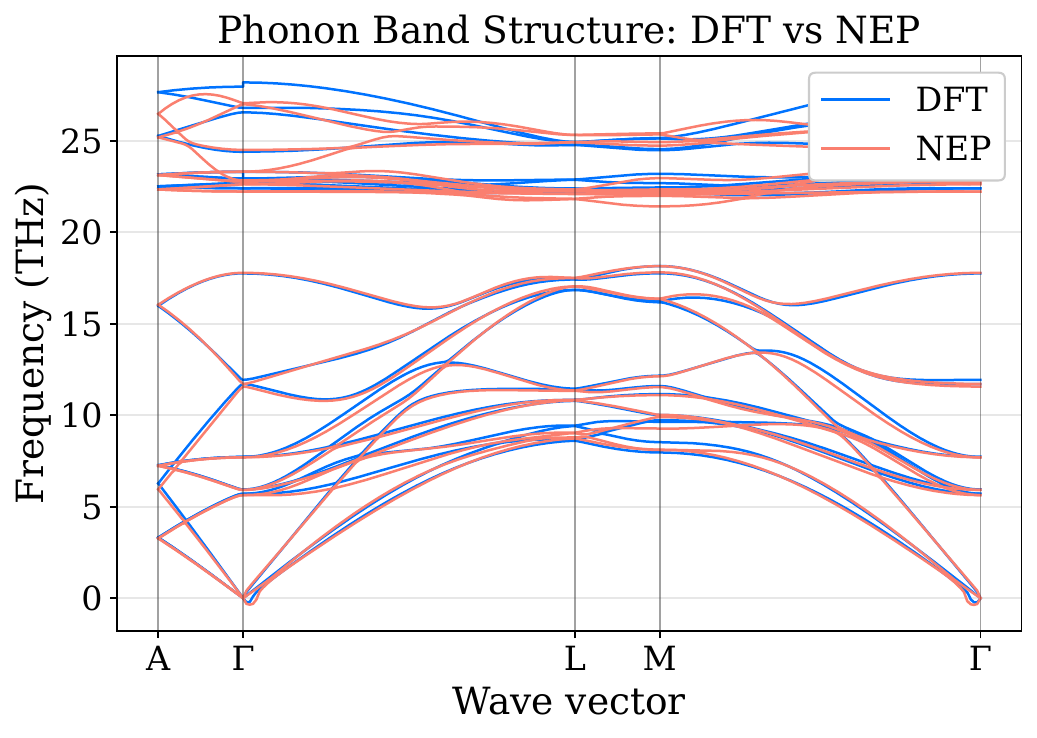}
        \caption{}
        \label{fig3:band_SiC}
    \end{subfigure}
    \begin{subfigure}[t]{0.45\linewidth}
        \centering
        \includegraphics[width=\linewidth]{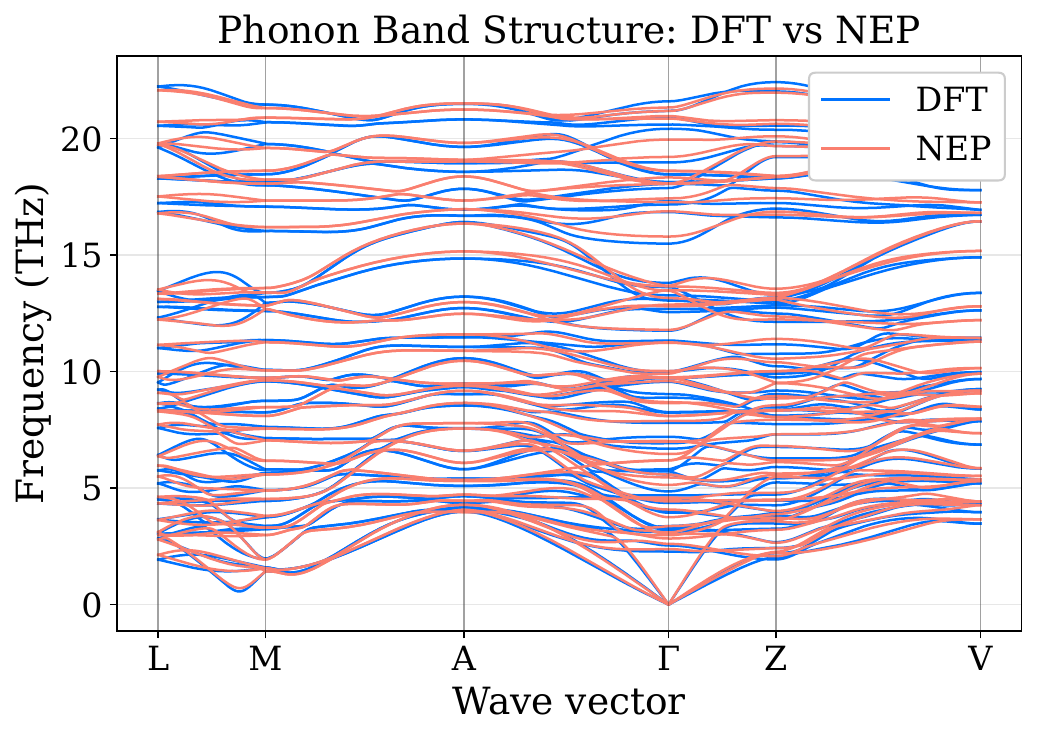}
        \caption{}
        \label{fig3:band_Ga2O3}
    \end{subfigure}
    \begin{subfigure}[t]{0.45\linewidth}
        \centering
        \includegraphics[width=\linewidth]{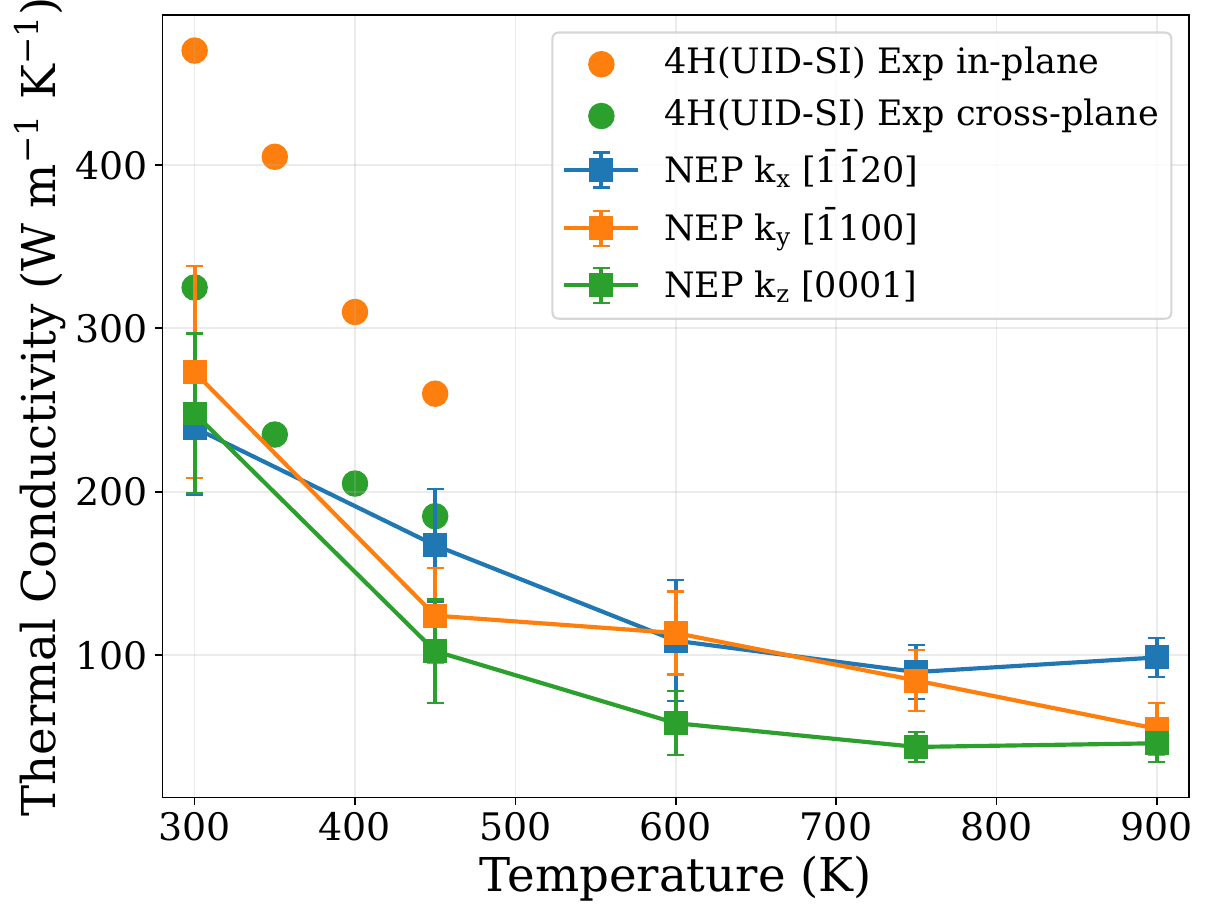}
        \caption{}
        \label{fig:kappa_SiC}
    \end{subfigure}
    \begin{subfigure}[t]{0.45\linewidth}
        \centering
        \includegraphics[width=\linewidth]{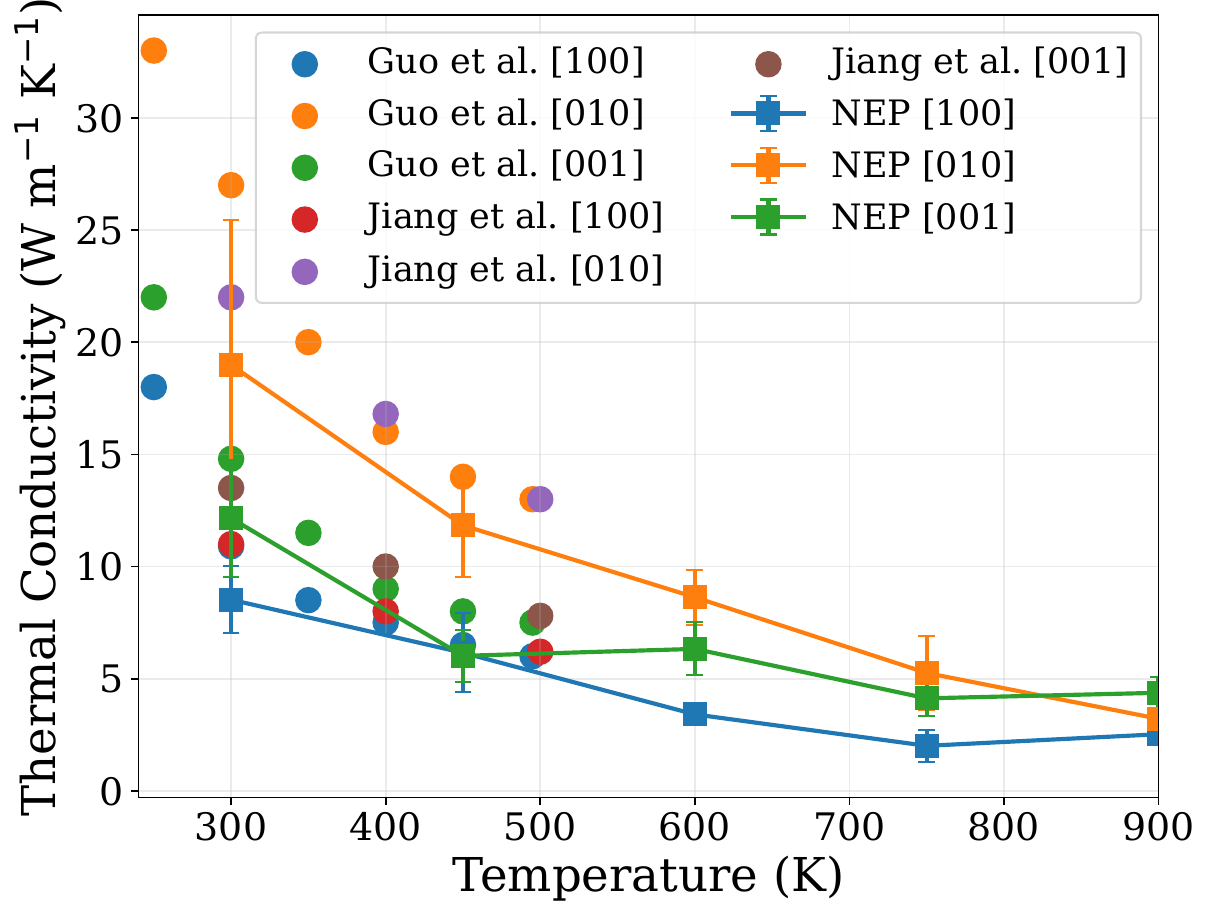}
        \caption{}
        \label{fig:kappa_Ga2O3}
    \end{subfigure}
    \caption{Bulk vibrational and thermal validation of the learned potential. Phonon dispersions of (a) SiC and (b) Ga$_2$O$_3$ predicted by DFT and NEP. Temperature-dependent thermal conductivity for (c) SiC \citep{Qian2017-SiC} and (d) Ga$_2$O$_3$ \citep{Guo2015-Ga, Jiang2018-Ga$_2$O$_3$} along different orientations from NEP (square markers) and pyhsical experiments (round markers)}
    \label{fig:kappa}
\end{figure}

The trained NEP is validated at both the supervised-fitting and transport-physics levels in Figures~\ref{fig2:nep-results} and~\ref{fig:kappa}. In Figure~\ref{fig2:nep-results}, the training losses decrease smoothly and converge stably, while the energy and force parity plots show near-diagonal agreement between NEP predictions and DFT labels. The test-set energy and force RMSEs are \(5.76\times10^{-3}\) eV and \(1.60\times10^{-1}\) eV/\AA, respectively, comparable to previously reported machine-learned interatomic potentials \citep{Fan2021-NEP,Zhao2023-mm}. These results show that a single feedforward NEP can represent oxide, carbide, and heterogeneous interfacial environments with near-DFT accuracy, while remaining far more efficient than direct first-principles calculations.

Figure~\ref{fig:kappa} further shows that the NEP reproduces the transport-relevant bulk physics of both constituents. For SiC and Ga$_2$O$_3$, the predicted phonon dispersions follow the DFT branches closely along the standard high-symmetry paths in the Brillouin zone, labeled by points such as \(\Gamma\), A, L, and M on the horizontal axis. In SiC, the main discrepancy appears near \(\Gamma\) around 25~THz, while in Ga$_2$O$_3$ several optical branches are slightly shifted, consistent with the fact that the DFT dispersions include non-analytical long-range dipole corrections, whereas local Machine-Learning Interatomic Potential (MLIPs) such as the present NEP describe interactions only within a finite cutoff. Importantly, at the lower frequency, the more dispersive branches that dominate heat transport are well reproduced.

The thermal-conductivity validation provides a complementary transport-level test. Figures~\ref{fig:kappa_SiC} and~\ref{fig:kappa_Ga2O3} show that the model preserves the key bulk trends of both materials: SiC remains much more conductive than Ga$_2$O$_3$, the conductivity of both phases decreases with temperature, and the principal anisotropic trends are maintained. Although the NEP underestimates absolute \(\kappa\), this is consistent with prior MLIP studies showing that force-fitting errors (i.e., small discrepancies between MLIP-predicted and DFT reference forces) can act as an additional perturbation and suppress thermal conductivity in MD \citep{Zhou2025-dd,Wu2024-po}. Taken together, these results indicate that the NEP captures the transport physics most relevant to the present study, making it suitable for mechanistic analysis of relative bulk and interfacial thermal-transport trends despite some underestimation of absolute \(\kappa\).

\subsection{Global TBC trends across length, temperature, and orientation}
\begin{figure}[!t]
\centering
\begin{minipage}[t]{0.55\linewidth}
    \centering
    \begin{subfigure}[t]{\linewidth}
        \centering
        \includegraphics[width=0.90\linewidth,trim=0 10 0 0, clip]{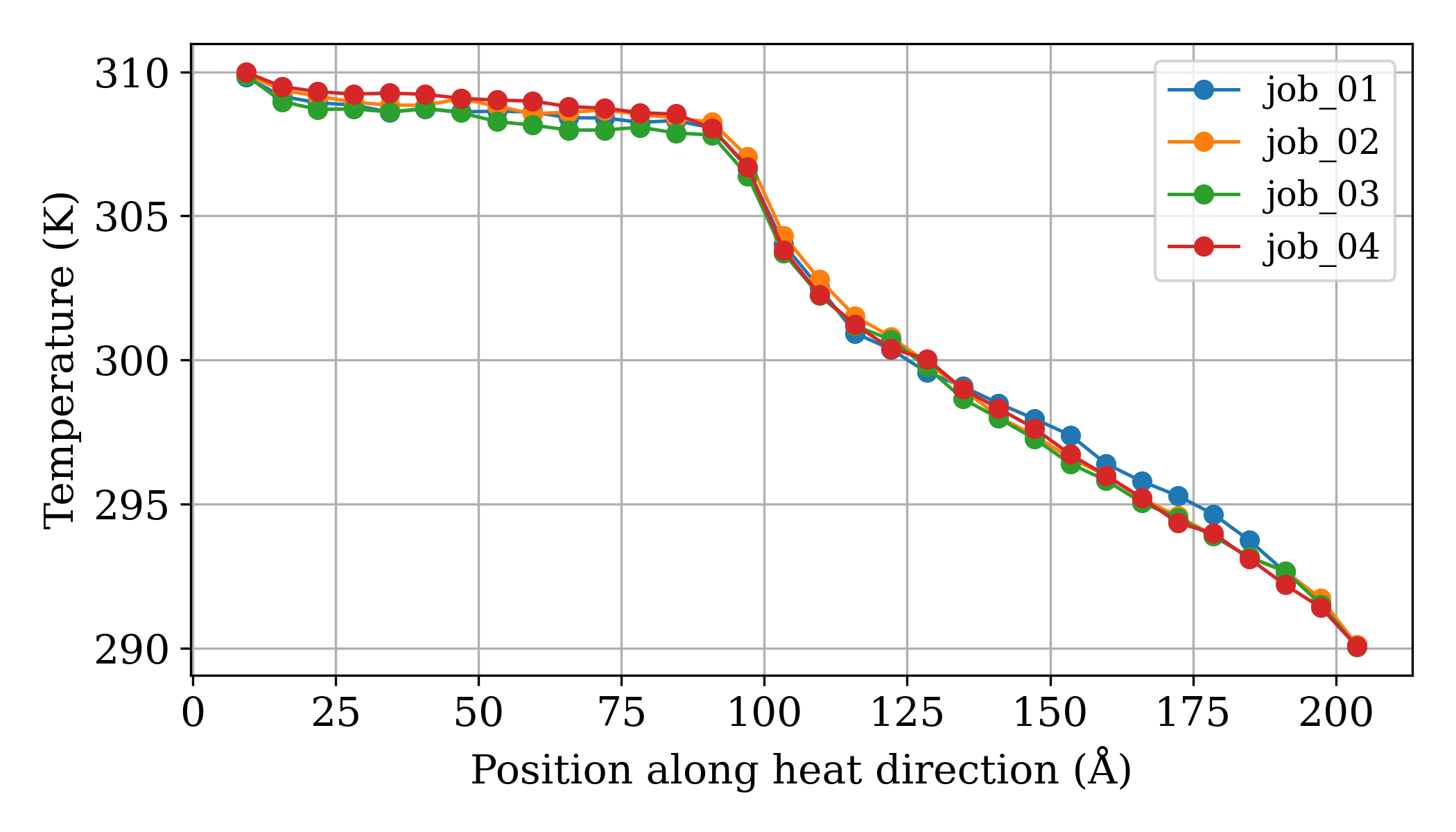}
        \caption{}
        \label{fig4:a}
    \end{subfigure}

    \vspace{0.8em}

    \vspace{1.0em}

    \hspace*{0.04\linewidth}
    \begin{subfigure}[t]{\linewidth}
        \centering
        \includegraphics[width=0.98\linewidth]{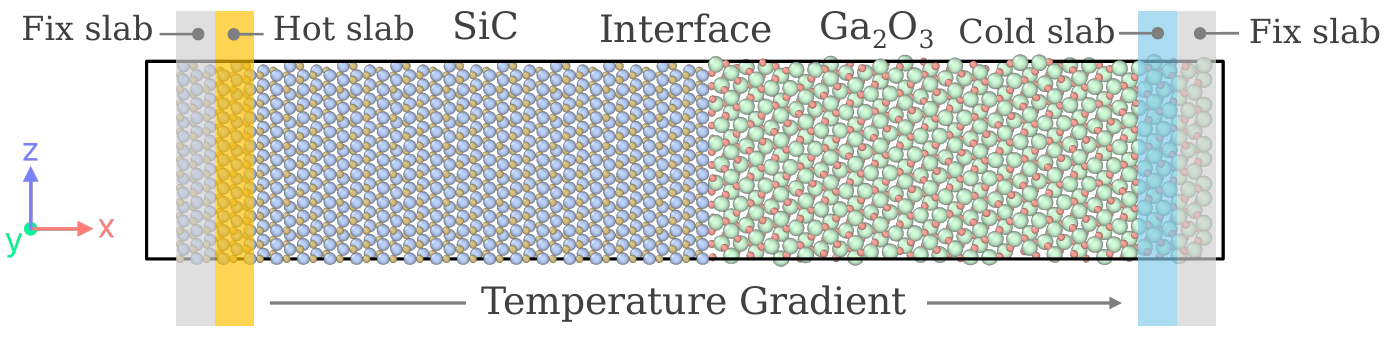}
        \caption{}
        \label{fig4:b}
    \end{subfigure}
\end{minipage}\hfill%
\begin{minipage}[t]{0.45\linewidth}
    \centering
\end{minipage}\hfill%
\begin{minipage}[t]{0.4\linewidth}
    \centering
    \begin{subfigure}[t]{\linewidth}
        \includegraphics[width=\linewidth,trim=0 5 0 0, clip]{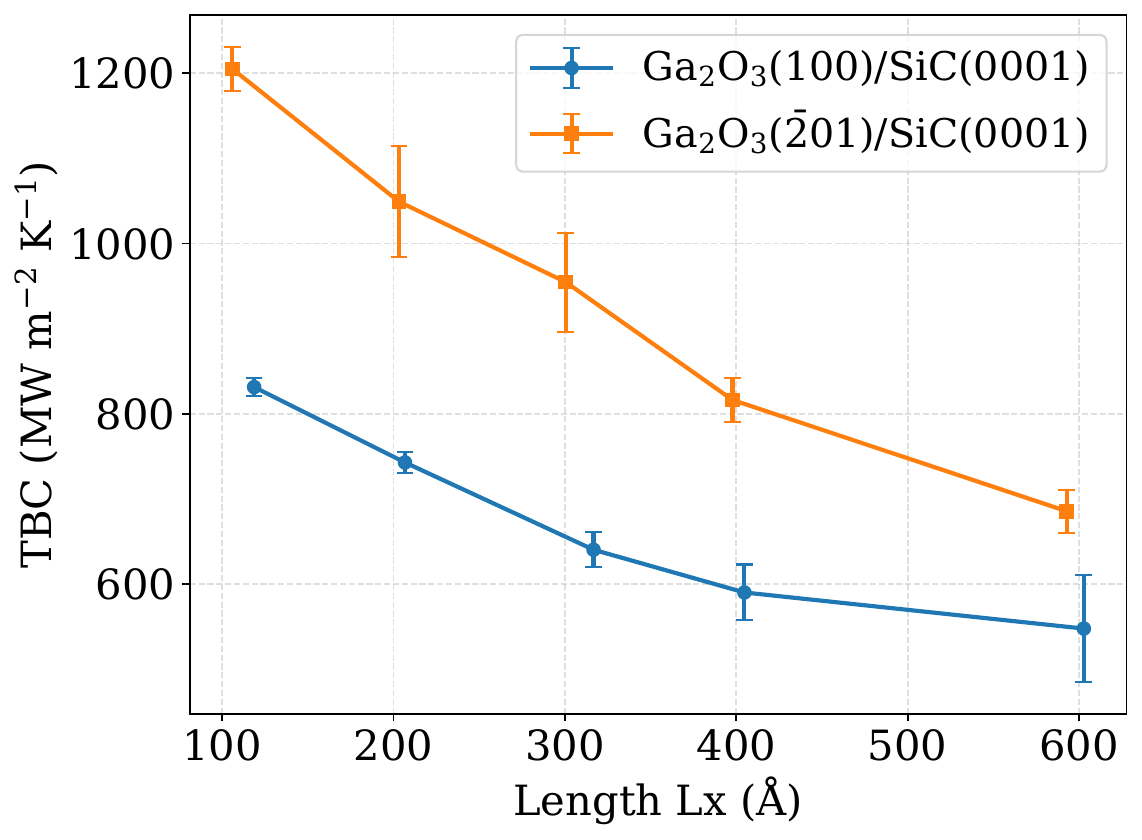}
        \caption{}
        \label{fig4:c}
    \end{subfigure}

    \vspace{0.5em}

    \begin{subfigure}[t]{\linewidth}
        \includegraphics[width=\linewidth,trim=0 5 0 0, clip]{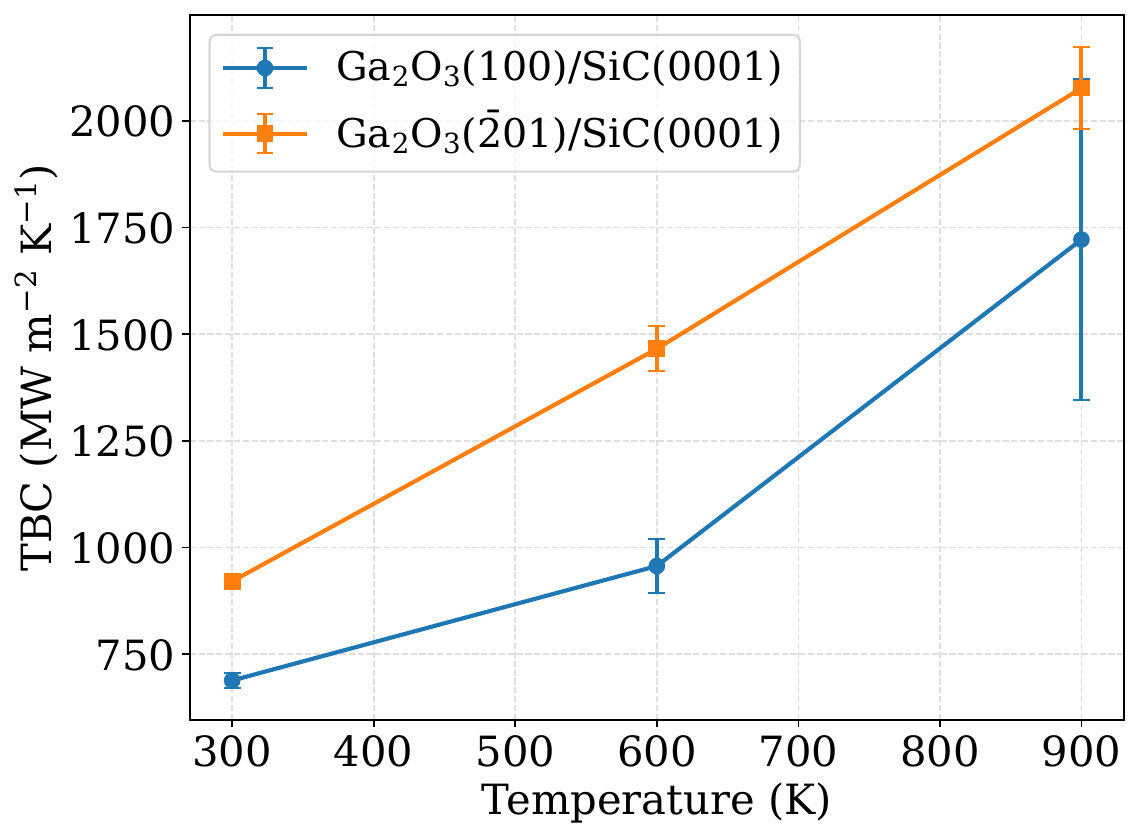}
        \caption{}
        \label{fig4:d}
    \end{subfigure}
\end{minipage}

\caption{Overall TBC trends from NEMD simulation. (a) Representative steady-state temperature profile. (b) Simulation cell and setup. (c) TBC versus transport length $L_x$. (d) TBC versus temperature. For both orientations, TBC decreases with increasing length and increases markedly with temperature.}
\label{fig:fig4}
\end{figure}

The central observation is that TBC depends systematically on transport length, temperature, and interface orientation, as shown in Figure~\ref{fig:fig4}. Panel (a) shows a representative steady-state temperature profile together with the simulation cell. In the NEMD setup, heat is continuously added to the hot slab and removed from the cold slab at opposite ends of the simulation bar, which establishes a steady heat flux across the interface. This produces an approximately linear temperature gradient within each slab and a clear temperature discontinuity at the interface. These features support the standard extraction of the interfacial temperature drop \(\Delta T_{\mathrm{int}}\) and heat flux \(J\), as described in Section~\ref{subsec:NEMD}. The error bars in panels (b) and (c) denote the standard deviation over four independent NEMD runs and therefore reflect run-to-run variability rather than fitting uncertainty.

Three robust trends emerge. First, at fixed temperature, the extracted TBC decreases monotonically with increasing simulation cell length along the transport (\(x\)) direction for both interface orientations, with the shortest cells showing the highest apparent conductance and the \SI{600}{\angstrom} cells the lowest. The decrease also weakens at larger lengths, suggesting progressive saturation. Second, at fixed length, the TBC increases markedly from \SI{300}{K} to \SI{900}{K}. Third, across all lengths and temperatures, the \((\bar{2}01)\) interface consistently outperforms the (100) interface, indicating intrinsically stronger interfacial heat transfer. At \SI{600}{\angstrom}, the predicted TBC falls in the range of 550--700 MW\,m\(^{-2}\)\,K\(^{-1}\), in good agreement with recent literature values of 500--950 MW\,m\(^{-2}\)\,K\(^{-1}\) \citep{Shen2025-NatCom, Zhang2026-IJHMT}. These trends motivate the mechanistic analysis below: the length dependence points to finite-size and carrier-supply effects, the temperature dependence suggests enhanced inelastic transport at elevated temperature, and the orientation dependence implies distinct structural and dynamical coupling at the two interfaces.

\subsection{Why TBC decreases with transport length: long-MFP carriers are attenuated before reaching the interface} \label{sec3.3}
\begin{figure}[t]
    \centering
    \begin{subfigure}[t]{0.45\linewidth}
        \centering
        \includegraphics[width=\linewidth]{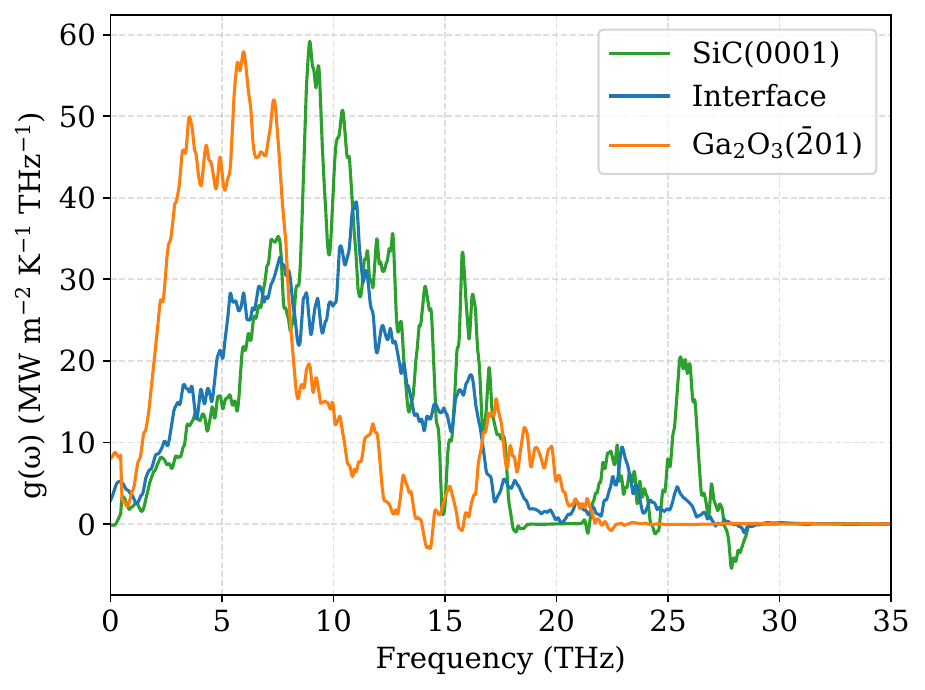}
        \caption{}
        \label{fig5a:shc}
    \end{subfigure}
    \begin{subfigure}[t]{0.45\linewidth}
        \centering
        \includegraphics[width=\linewidth]{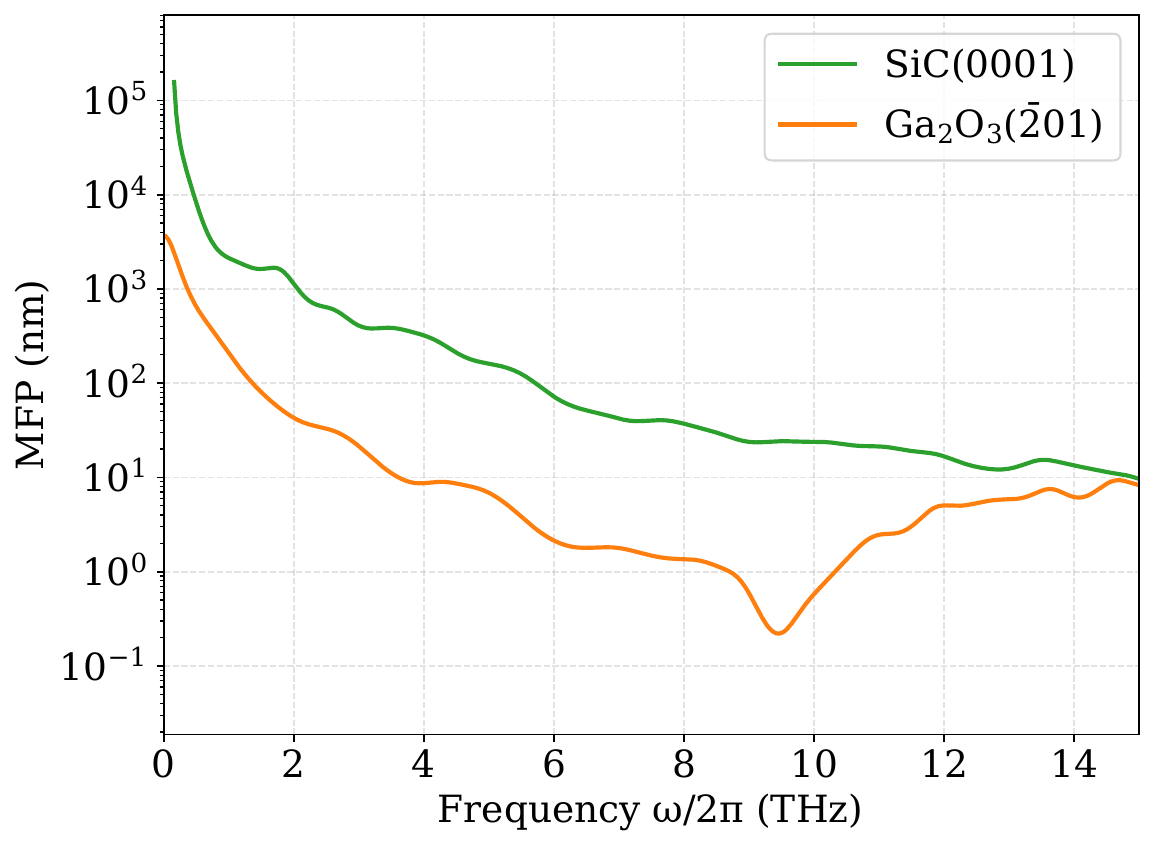}
        \caption{}
        \label{fig5b:mfp}
    \end{subfigure}
    \caption{Mechanistic interpretation of the length-dependent TBC. (a) Representative frequency-resolved spectral conductance for the SiC side, interface region, and Ga$_2$O$_3$ side. (b) Phonon mean free path spectra of SiC and Ga$_2$O$_3$. Those panels show that the lower TBC with increasing length arises mainly from attenuation of long-mean-free-path carriers before they reach the interface.}
    \label{fig5:MFP}
\end{figure}

Figure~\ref{fig5:MFP} explains the length dependence of TBC from two complementary perspectives. Panel (a) shows the spectral conductance \(g(\omega)\), which measures the frequency-resolved contribution to interfacial heat transfer. The green, blue, and orange curves correspond to 5~\AA-wide slabs sampled in the SiC region, the interface region, and the Ga$_2$O$_3$ region, respectively. The interfacial conductance is concentrated mainly in the low- and intermediate-frequency range, roughly 5--15~THz, where vibrational modes from the two materials overlap and couple efficiently across the interface.

Panel (b) shows the phonon mean free path (MFP) spectra of bulk SiC and Ga$_2$O$_3$, where Ga$_2$O$_3$($\bar{2}$01) orientation is taken for an example. Phonons are the main heat-carrying lattice vibrations in these crystals, and the MFP characterizes how far they travel before being scattered inside each slab. The MFP here is calculated based on the ratio between diffusive conductivity and the ballistic conductance per unit area \citep{Dong2024-tt}. A key observation is that many low-frequency SiC phonons have much longer mean free paths, on the order of tens to hundreds of nanometers, than those in Ga$_2$O$_3$. For bar lengths on the order of \(10^2\) nm (\(10^3\) \AA), these long-MFP carriers can reach the interface with limited internal scattering and efficiently supply the transmissive spectral window identified in panel (a). As the transport length increases, however, they are progressively scattered within the slabs before arriving at the interface, which reduces the effective heat flux available for interfacial transfer. Therefore, the observed drop in TBC with increasing length does not imply that the interface itself becomes intrinsically less transmissive; rather, it reflects reduced ballistic or quasi-ballistic carrier supply as the bar length approaches or exceeds the relevant MFPs.

Although macroscopic heat conduction is typically measured at much larger length scales, experimentally deposited Ga$_2$O$_3$ layers on SiC are often only 60--120~nm thick \citep{Shen2025-NatCom,Nepal2020-TBC-exp}. At these dimensions, size effects and quasi-ballistic phonon transport can still be significant. Therefore, the present analysis is not only a simulation artifact of nanoscale cells, but also provides a physically relevant reference for understanding the thickness dependence of TBC in experimentally accessible structures.

\begin{figure}[t]
    \centering
    \begin{subfigure}[t]{0.45\linewidth}
        \centering
        \includegraphics[width=\linewidth,trim=0 5 0 5, clip]{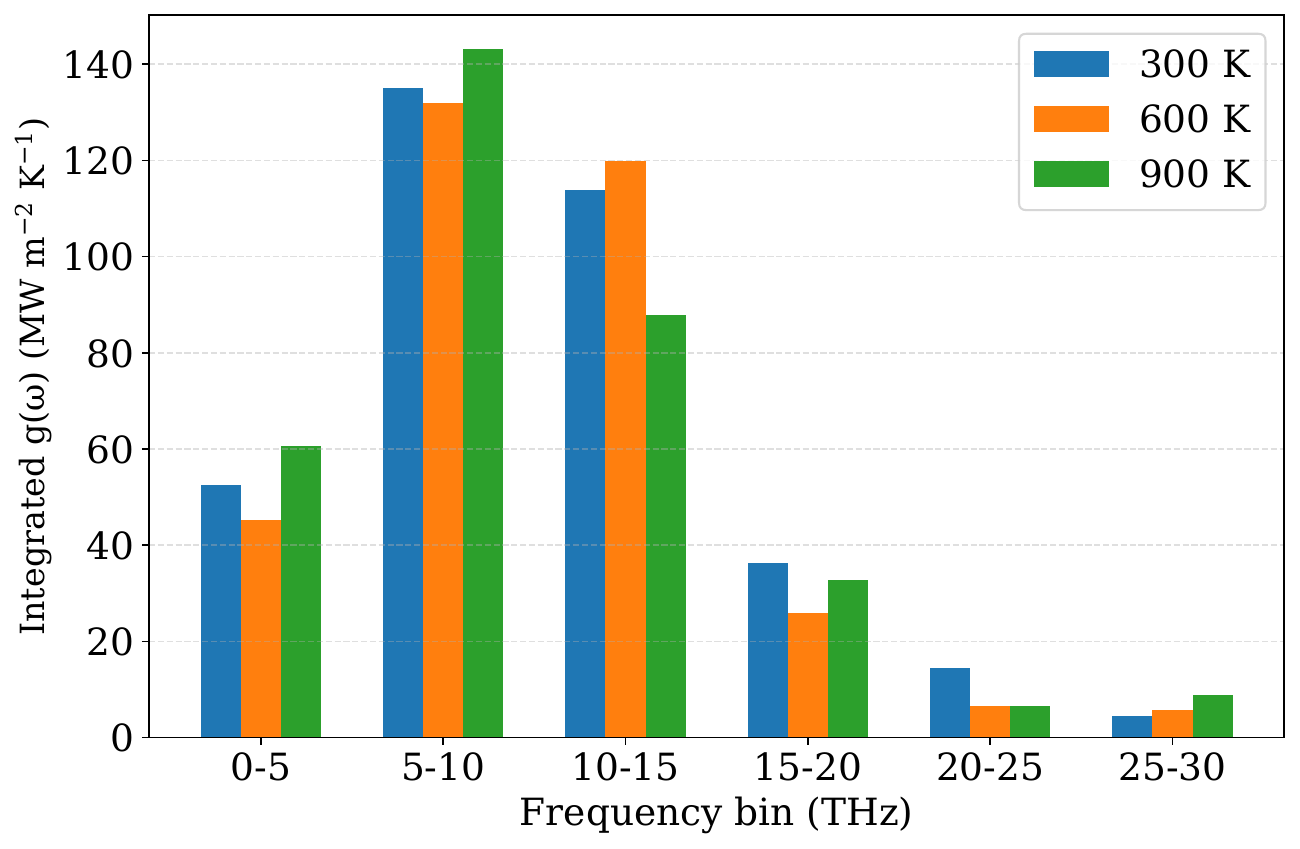}
        \caption{}
        \label{fig6a:}
    \end{subfigure}
    \begin{subfigure}[t]{0.45\linewidth}
        \centering
        \includegraphics[width=\linewidth,trim=0 5 0 0, clip]{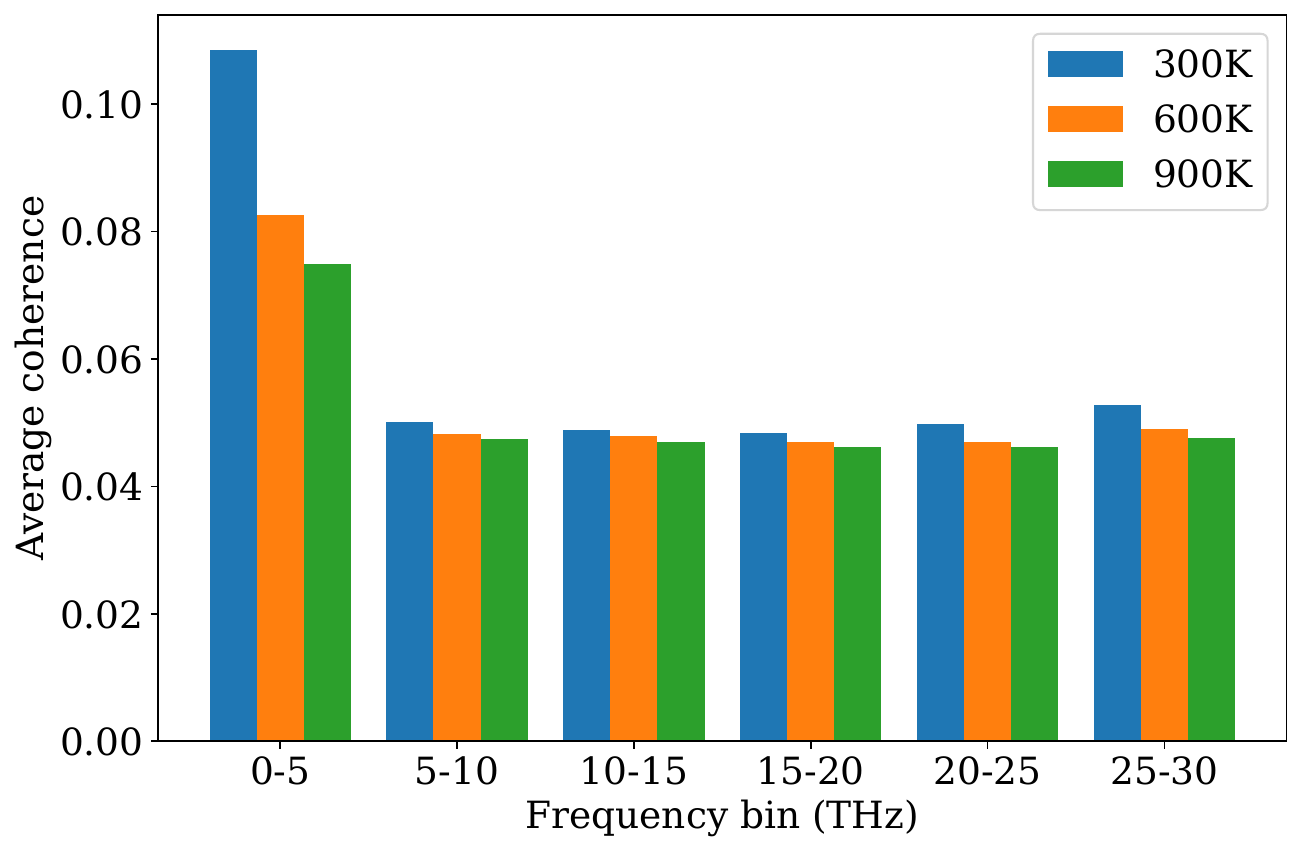}
        \caption{}
        \label{fig6b:}
    \end{subfigure}
    \begin{subfigure}[t]{0.45\linewidth}
        \centering
        \includegraphics[width=\linewidth,trim=0 5 0 0, clip]{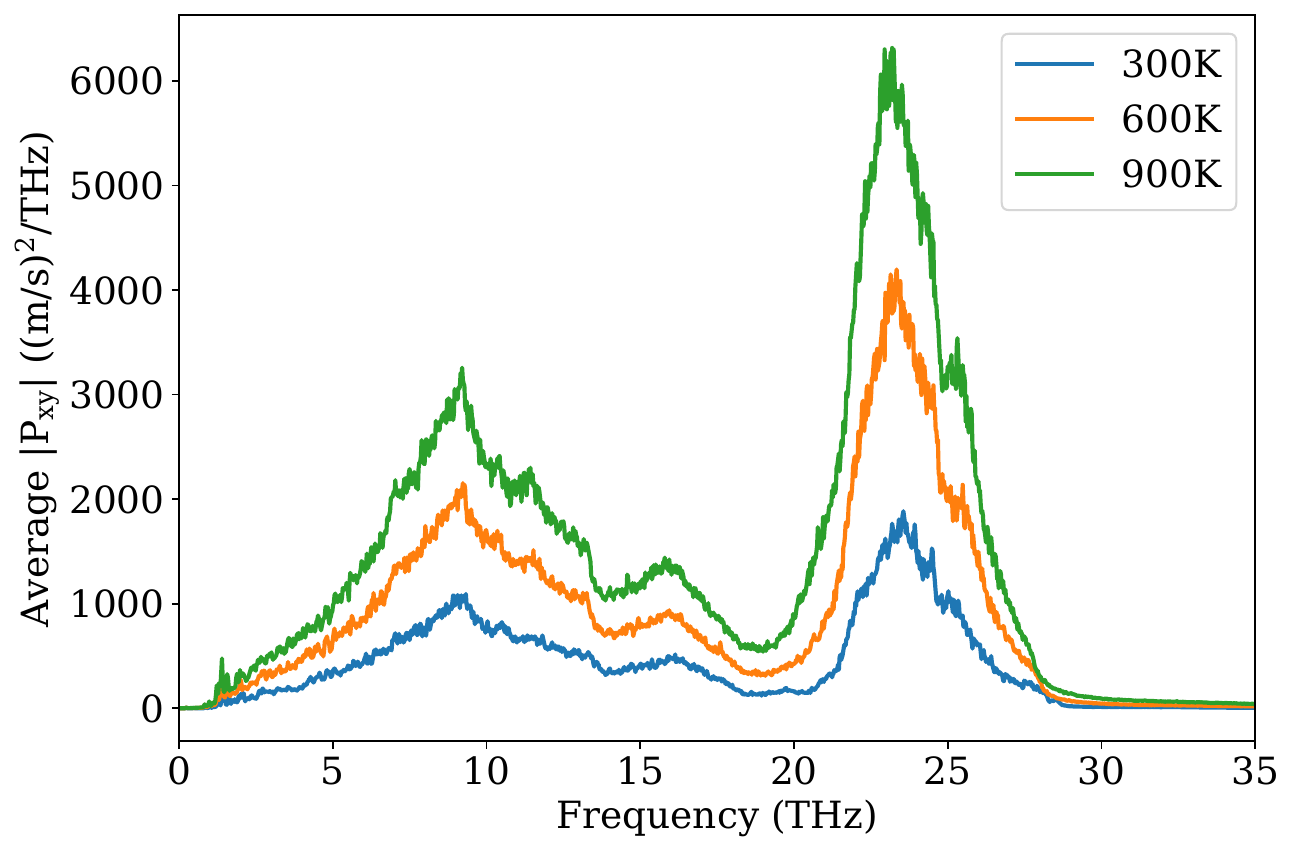}
        \caption{}
        \label{fig6c:}
    \end{subfigure}
    \caption{Mechanistic interpretation of the temperature-dependent TBC. (a) Frequency-binned spectral conductance. (b) Frequency-binned cross-interface coherence. (c) Cross-spectral density at the interface. Those panels show broadly unchanged transport channels, reduced coherence, and stronger interfacial dynamical coupling at higher temperature.}
    \label{fig6:TBC_temp_explanation}
\end{figure}

\subsection{Why TBC increases with temperature: stronger inelastic interfacial coupling} \label{sec3.4}
The physical mechanism of the TBC dependence on temperature is well explained in Figure~\ref{fig6:TBC_temp_explanation}. Panel (a) shows the binned spectral conductance, which measures how different frequency ranges contribute to interfacial heat transfer. The dominant transport window remains broadly similar from \SI{300}{K} to \SI{900}{K}, with the main contributions still concentrated in the mid-frequency range. This indicates that the TBC increase is not primarily caused by the opening of entirely new transport channels at the high temperatures.

Panel (b) shows the cross-interface coherence defined in Appendix~\ref{app:equations}, which quantifies how phase-preserving the interfacial transport is at each frequency. The coherence decreases with increasing temperature, especially in the low-frequency range, indicating that transport becomes less coherent at elevated temperature. This trend proposes a picture in which the TBC enhancement is not dominated by stronger coherent elastic transmission alone.

Panel (c) helps clarify the temperature mechanism. The cross-spectral density measures the frequency-resolved dynamical correlation between atomic motions on the two sides of the interface; larger values indicate stronger interfacial vibrational interaction, even if that interaction is not fully phase-preserving. As temperature increases, the cross-spectral density rises markedly over the main transport window, showing stronger interfacial coupling. Together with Appendix~\ref{app:coupling}, which shows that both diagonal and off-diagonal coupling increase with temperature, this indicates that elastic and inelastic exchange are both enhanced. However, because the cross-interface coherence simultaneously decreases, the TBC increase is more consistently attributed to a relatively stronger growth of incoherent and inelastic transfer within a broadly unchanged transport window.

This trend is also relevant for device operation. As temperature rises, the increase in TBC helps improve interfacial heat dissipation rather than further bottlenecking it. These results suggest that interface modifications that enhance anharmonic cross-interface coupling could improve heat dissipation at elevated temperatures.

\subsection{Why the \texorpdfstring{$(\bar{2}01)$}{(2bar01)} interface outperforms the (100) interface: stronger bonding and larger cross-interface coupling} \label{sec3.5}

\begin{figure}[t]
    \centering
    \begin{subfigure}[t]{0.42\linewidth}
        \centering
        \includegraphics[width=\linewidth,trim=0 5 0 5, clip]{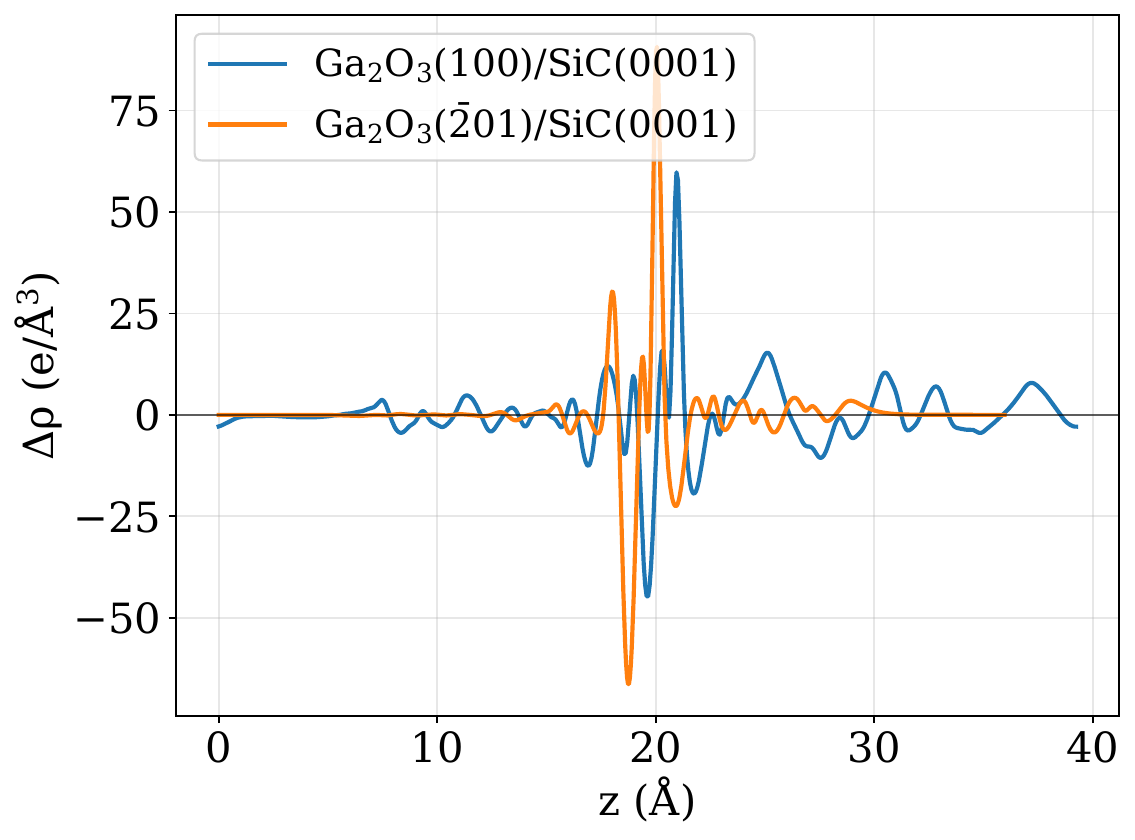}
        \caption{}
        \label{fig7a}
    \end{subfigure}
    \begin{subfigure}[t]{0.53\linewidth}
        \centering
        \includegraphics[width=\linewidth,trim=0 5 0 0, clip]{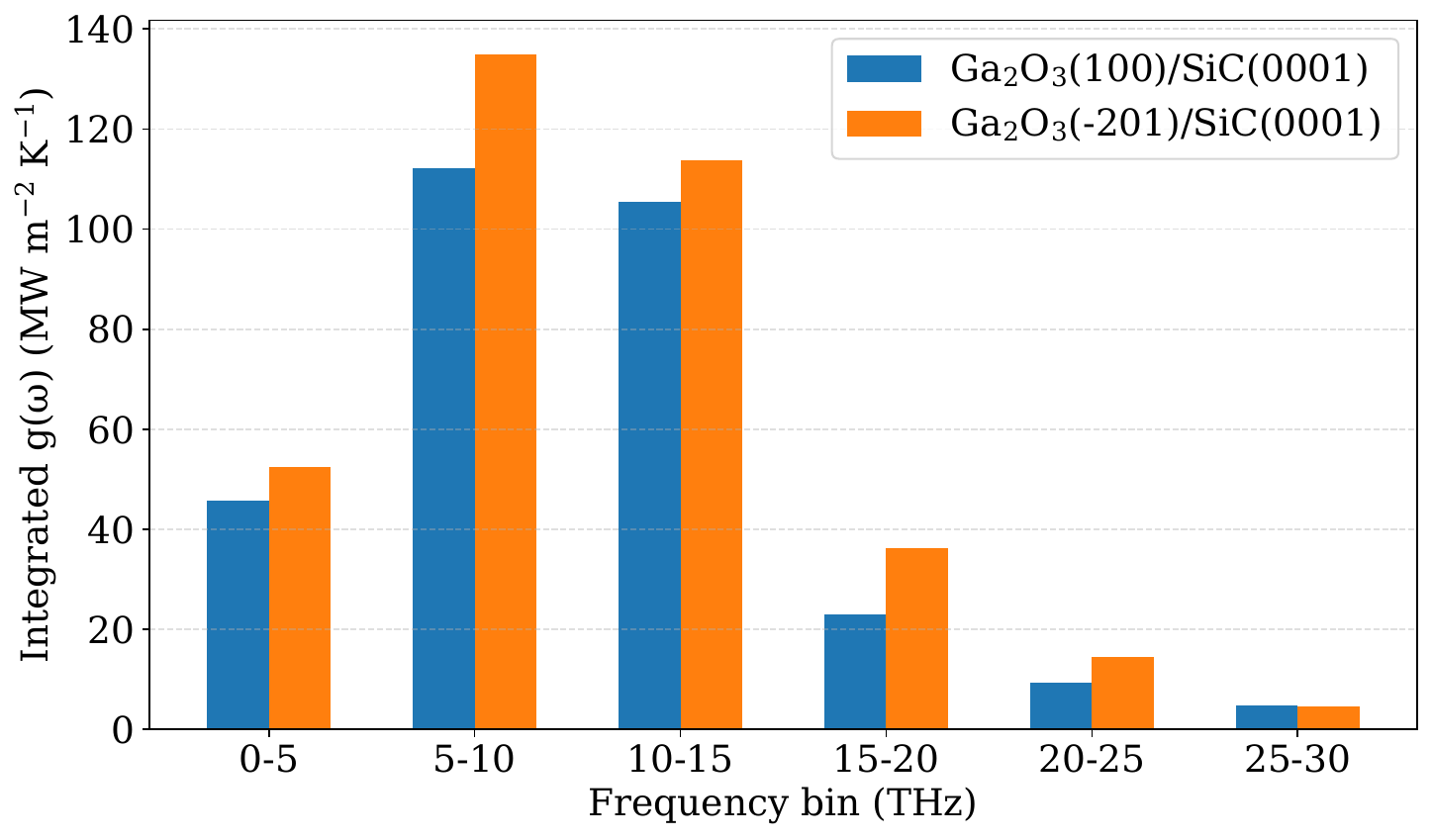}
        \caption{}
        \label{fig7b}
    \end{subfigure}
    \caption{Orientation-dependent interfacial diagnostics. (a) Planar-averaged charge-density-difference profile \(\Delta \rho(z)\), showing distinct interfacial charge redistribution for the two orientations. (b) Frequency-binned integrated spectral conductance, showing stronger vibrational transmission for the $(\bar{2}01)$ interface, particularly in the main 5--15~THz transport range.}
    \label{fig7:orientation_mech}
\end{figure}

To understand the orientation dependence of TBC, we examine static and dynamical descriptors of bonding, vibrational transmission, and force coupling based on DFT and NEP-based MD, as summarized in Figure~\ref{fig7:orientation_mech} and Table~\ref{tab:orientation}. Panel (a) shows the planar-averaged charge-density difference, which measures how electrons redistribute when the two slabs are brought into contact relative to the isolated constituents. This quantity is important because interfacial heat transfer is ultimately controlled by bonding and force transmission across the interface, and the charge redistribution profile provides a direct electronic-structure signature of how the interface bonds are formed. The two orientations show clearly different redistribution patterns, indicating distinct local bonding topologies rather than a simple geometric rotation of the same interface.

Panel (b) shows the integrated spectral conductance in different frequency bins. The $(\bar{2}01)$/SiC(0001) interface consistently exhibits larger conductance in the most important 5--15~THz range, which is the main transport window identified earlier, and it also shows larger contributions in several higher-frequency bins. This indicates that the superior TBC of the $(\bar{2}01)$ interface is not due to a single narrow spectral feature, but to broadly stronger vibrational transmission across the frequencies that matter most for interfacial heat flow.

The structural and energetic metrics in Table~\ref{tab:orientation} further reveal the static origin of this orientation dependence, all the related definitions could be found in Appendix~\ref{app:equations}. Relative to the (100) interface, the $(\bar{2}01)$ interface shows larger total charge transfer, larger charge transfer per unit area, and a more negative binding energy, all of which indicate stronger interfacial adhesion. Most importantly, all three normalized interfacial force-coupling metrics, \(K_{\mathrm{cross}}/A\), \(K_{nn}/A\), and \(K_{tt}/A\), are larger for the $(\bar{2}01)$ interface. These metrics quantify the strength of the harmonic cross-interface force constants, i.e., how effectively atomic displacements on one side generate restoring forces on the other.

It is worth noting that temperature and orientation affect TBC through different mechanisms, although both lead to stronger interfacial heat transfer. Changing the orientation mainly alters the intrinsic nature of the interface: the $(\bar{2}01)$ configuration forms stronger interfacial bonding and larger harmonic force coupling, analogous to constructing a better bridge with a more robust structure. Increasing temperature, by contrast, does not fundamentally change the bridge itself, but enhances dynamic vibrational interaction and anharmonic or inelastic energy exchange across the existing interface. Thus, orientation improves the intrinsic transmission capability of the interface, whereas temperature mainly improves the transport efficiency across that interface.

All of these measures, from electronic structure to heat conductance, support the same conclusion: the higher TBC of Ga$_2$O$_3$$(\bar{2}01)$/SiC(0001) arises from a more favorable atomic arrangement that strengthens interfacial bonding and cross-interface vibrational coupling. Engineering the interface orientation is therefore a meaningful design variable for device-level thermal transport. In the present system, changing the interface from (100) to $(\bar{2}01)$ increases the TBC by roughly 20\%, demonstrating that differences in local symmetry, bonding geometry, and charge redistribution can substantially modify heat-transfer performance.

\begin{table}[t]
    \centering
    \small
    \setlength{\tabcolsep}{5pt}
    \caption{Key orientation-dependent interfacial metrics. The $(\bar{2}01)$ interface consistently shows stronger bonding and coupling than the (100) interface.}
    
    \label{tab:orientation}
    \begin{tabular}{p{0.38\columnwidth} c c c}
        \toprule
        Metric & Unit & Ga$_2$O$_3$(100)/SiC(0001) & Ga$_2$O$_3$($\bar{2}01$)/SiC(0001) \\
        \midrule
        Interface area & \AA$^2$ & 34.8 & 49.1 \\
        Charge transfer per area & $e$/\AA$^2$ & 0.0823 & 0.0921 \\
        Interface binding energy & eV/\AA$^2$ & -0.671 & -0.715 \\
        Average interfacial Si--O bond length & \AA & 1.66 & 1.66 \\
        $K_{\mathrm{cross}}/A$ & --- & 4.83 & 4.90 \\
        $K_{nn}/A$ & --- & 4.10 & 4.17 \\
        $K_{tt}/A$ & --- & 0.974 & 1.04 \\
        \bottomrule
    \end{tabular}
\end{table}

    

Results from Sections~\ref{sec3.3}--\ref{sec3.5} show that even a compact feedforward neural network potential can support physics-grounded understanding of thermal boundary conductance. The length and temperature mechanisms are resolved directly from NEP-based MD transport analysis, while the orientation dependence is further supported by complementary electronic-structure and bonding evidence. Thus, the model serves not only as an efficient large-scale predictor, but also as a tool for identifying the physical factors governing interfacial heat-transfer trends.

\section{Conclusion, Limitations, and Future Work}

We developed a unified Ga--O--Si--C neuroevolution potential to study thermal boundary conductance at Ga$_2$O$_3$/SiC interfaces. The model achieves low supervised errors, reproduces bulk phonons and thermal-conductivity trends, and is sufficiently transferable for nonequilibrium transport simulations across oxide/carbide interfaces. Using this potential, we find that TBC decreases with transport length, increases strongly with temperature, and is consistently higher for Ga$_2$O$_3$($\bar{2}01$)/SiC(0001) than for Ga$_2$O$_3$(100)/SiC(0001), reaching up to about 700 MW\,m\(^{-2}\)\,K\(^{-1}\). These results provide an atomistic basis for thermal management in Ga$_2$O$_3$/SiC-based power electronics, where interface-limited heat dissipation can strongly affect performance and reliability.

The transport mechanisms are resolved through physically interpretable analysis of the NEP-based simulations. The length dependence mainly reflects carrier supply, as long-mean-free-path modes in shorter slabs reach the interface with less internal scattering and produce higher apparent TBC. The temperature dependence arises not from a major shift of the transport window, but from stronger incoherent and anharmonic cross-interface exchange at elevated temperature despite reduced coherence. The orientation dependence is governed by interfacial structure, with stronger bonding, larger charge transfer, and larger cross-interface force coupling giving higher conductance for the $(\bar{2}01)$ interface. Overall, this work shows that a single transferable machine-learning potential can support both large-scale transport prediction and physics-grounded mechanistic understanding of interfacial heat flow in semiconductor heterostructures.

Several limitations remain. The present study considers ideal, atomically ordered interfaces, whereas real devices may contain roughness, intermixing, defects, and processing-induced disorder. The transport analysis is classical, so quantum nuclear effects are not explicitly included, and no generally accepted rigorous quantum-correction framework exists for interfacial conductance.  Future work should therefore address more realistic non-ideal interfaces and the generality of the physical trends identified here beyond Ga$_2$O$_3$/SiC.

\bibliographystyle{unsrtnat}
\bibliography{references}

\appendix

\section{List of Abbreviations}
\label{app:abbrev}

This section summarizes the abbreviations used throughout this work, together with their full names and brief explanations.

\begin{itemize}
\small
\item \textbf{AIMD -- Ab Initio Molecular Dynamics} \\
A simulation method in which atomic motion is computed using first-principles quantum mechanical calculations. It is used to generate accurate reference data for model training.

\item \textbf{DFT -- Density Functional Theory} \\
A quantum mechanical approach for computing the electronic structure of materials. It provides high-accuracy reference energies, forces, and stresses.

\item \textbf{MD -- Molecular Dynamics} \\
A computational technique that simulates the time evolution of atomic systems using classical mechanics.

\item \textbf{EMD -- Equilibrium Molecular Dynamics} \\
A method for computing transport properties under equilibrium conditions, typically using Green--Kubo relations.

\item \textbf{HNEMD -- Homogeneous Nonequilibrium Molecular Dynamics} \\
A simulation approach that applies an external driving field to induce a heat current and evaluate thermal conductivity efficiently.

\item \textbf{NEMD -- Nonequilibrium Molecular Dynamics} \\
A method that imposes a temperature gradient to directly calculate heat transport properties such as thermal boundary conductance.

\item \textbf{MLIP -- Machine-Learning Interatomic Potential} \\
A data-driven model that approximates atomic interactions using machine learning, combining high accuracy with computational efficiency.

\item \textbf{NEP -- Neuroevolution Potential} \\
A type of machine-learned interatomic potential based on a feedforward neural network optimized through evolutionary strategies.

\item \textbf{NN -- Neural Network} \\
A computational model composed of interconnected neurons that learns patterns from data; used here to map atomic environments to energies.

\item \textbf{TBC -- Thermal Boundary Conductance} \\
A measure of the efficiency of heat transfer across an interface between two materials.

\item \textbf{IFC -- Interatomic Force Constant} \\
A quantity describing the strength of atomic interactions under small displacements, used to analyze vibrational and thermal properties.

\item \textbf{MFP -- Mean Free Path} \\
The average distance a phonon travels before scattering, which determines heat transport behavior.

\item \textbf{SHC -- Spectral Heat Current} \\
A frequency-resolved measure of heat transfer used to analyze contributions from different vibrational modes.

\item \textbf{RMSE -- Root Mean Square Error} \\
A statistical metric used to quantify the difference between predicted and reference values.

\item \textbf{GPU -- Graphics Processing Unit} \\
A hardware accelerator used for efficient parallel computation in simulations and neural network training.

\item \textbf{PBE -- Perdew--Burke--Ernzerhof Functional} \\
A commonly used exchange-correlation functional within density functional theory.

\end{itemize}

\section{Loss function}
\label{app:nep-loss}
The NEP model is trained by minimizing a weighted multi-objective loss that combines errors in energies, atomic forces, and virials:
\begin{equation}
\begin{aligned}
\mathcal{L} =\;&
\lambda_E
\left[
\frac{1}{N_{\mathrm{str}}}
\sum_{n=1}^{N_{\mathrm{str}}}
\left(
U_n^{\mathrm{NEP}}-U_n^{\mathrm{DFT}}
\right)^2
\right]^{1/2}
+
\lambda_F
\left[
\frac{1}{3N}
\sum_{i=1}^{N}
\left\|
\mathbf{F}_i^{\mathrm{NEP}}-\mathbf{F}_i^{\mathrm{DFT}}
\right\|^2
\right]^{1/2} \\
&+
\lambda_V
\left[
\frac{1}{6N_{\mathrm{str}}}
\sum_{n=1}^{N_{\mathrm{str}}}
\sum_{\mu\nu}
\left(
W_{\mu\nu,n}^{\mathrm{NEP}}-W_{\mu\nu,n}^{\mathrm{DFT}}
\right)^2
\right]^{1/2},
\end{aligned}
\end{equation}
with additional regularization terms included in the full implementation. Here, \(N_{\mathrm{str}}\) is the number of structures in a batch, \(N\) is the total number of atoms, \(U\) denotes per-structure energies, \(\mathbf{F}_i\) the atomic forces, and \(W_{\mu\nu}\) the virial components. This multi-objective loss constrains not only the energy landscape, but also its first derivatives and stress response, which are all essential for reliable phonon and interfacial transport simulations. 

\section{Training parameters}
\label{app:nep-training}

\begin{table}[H]
\centering
\caption{Summary of the NEP training dataset and training hyperparameters. Interface structures include multiple Ga$_2$O$_3$/SiC slab models with different orientations, registries, distorted configurations, and thermally sampled snapshots. H atoms are included only as passive surface passivants in selected configurations.}
\label{tab:nep_training_setup}
\small
\begin{tabular}{lll}
\toprule
\textbf{Category} & \textbf{Item} & \textbf{Value} \\
\midrule
\multirow{7}{*}{Dataset}
& SiC structures & 5795 \\
& Ga$_2$O$_3$ structures & 6852 \\
& Interface structures & 16609 \\
& Total structures & 29256 \\
& Train / val / test & 21065 / 5266 / 2925 \\
& Split ratio & 72.0\% / 18.0\% / 10.0\% \\
\midrule
\multirow{11}{*}{NEP setup}
& Type map & Ga, O, Si, C, H \\
& Number of generations & 500000 \\
& Batch size & 1000 \\
& \(n_{\max}\) & 10, 8 \\
& Basis size & 12, 10 \\
& Hidden neurons & 96 \\
& Population size & 24 \\
& Cutoff radii & 9.0, 6.0~\AA \\
& \(\lambda_1\) & 0.05 \\
& \(\lambda_2\) & 0.05 \\
& \(\lambda_E, \lambda_F, \lambda_V\) & 1, 2, 0.1 \\
\bottomrule
\end{tabular}
\end{table}

\section{Simulation details}
\label{app:sim}
\subsection{Dataset Construction and Potential Validation}
Most training configurations were generated by \textit{ab initio} molecular dynamics (AIMD) followed by single-point density functional theory (DFT) calculations, with a smaller set of larger-supercell snapshots first sampled using the universal M3GNet potential and then refined by single-step DFT calculations \citep{Chen2022}. For each interface orientation, two supercell sizes were constructed. For the Ga$_2$O$_3$(100)/SiC(0001) interface, the models contain 106 and 168 atoms, whereas for the Ga$_2$O$_3$($\bar{2}01$)/SiC(0001) interface, the corresponding models contain 69 and 106 atoms. The lattice mismatch of these interface structures is in the range of approximately 2\%--5\%. First-principles calculations were performed with VASP 6.4 \citep{Hafner2008} using the generalized gradient approximation with the PBE functional \citep{Ernzerhof1999}, a plane-wave cutoff of 350~eV, and an electronic convergence threshold of $10^{-4}$~eV. Because the structures have varied cell sizes, the Brillouin zone was sampled using a uniform reciprocal-space resolution of 0.3~\AA$^{-1}$ rather than a fixed Monkhorst--Pack \textit{k}-point mesh. The resulting dataset spans bulk and interface-containing structures over a wide range of thermal distortions, including AIMD snapshots sampled during heating from near 0~K to 1000~K over 4000 MD steps, together with a limited number of high-temperature ($>2000$~K) melted configurations to improve robustness and transferability. These reference calculations provide the energies, forces, and virials used to train the unified neural network potential for both near-equilibrium lattice dynamics and distorted local environments relevant to interfacial transport.

\subsection{MD simulations}
To validate the bulk transport fidelity of the trained potential, we performed equilibrium molecular dynamics (EMD) simulations in LAMMPS \citep{Thompson2022} and extracted thermal conductivity from the Green--Kubo formalism \citep{Green1954,Kubo1957}. For bulk SiC, we used a $16\times16\times5$ conventional supercell containing 10240 atoms, while for bulk $\beta$-Ga$_2$O$_3$ we used a $4\times16\times8$ conventional supercell containing 10240 atoms, so that the simulation boxes were sufficiently large for size-converged EMD calculations. Periodic boundary conditions were applied in all three directions. Each system was equilibrated for 200~ps in the isothermal--isobaric ensemble and then propagated for 2~ns in the microcanonical ensemble using a 1.0~fs time step. Thermal conductivity values from 300 to 900~K were averaged over 10 independent trajectories.

We also performed HNEMD simulations to evaluate bulk thermal conductivity more efficiently and to cross-check the transport trends obtained from EMD. In HNEMD, an external driving field is applied to induce a homogeneous heat current throughout the periodic simulation cell, and the thermal conductivity is obtained from the linear-response relation between the induced heat current and the applied field. The HNEMD simulations were carried out with periodic boundary conditions in all three directions and a time step of 1.0~fs. For each target temperature \(T\), the system was first equilibrated for 1~ns in the isothermal--isobaric ensemble and then evolved for 5~ns in the canonical ensemble for data collection. The external driving field was applied along the transport direction with magnitude \(2\times10^{-5}\)~\AA\(^{-1}\). The conductivity at each temperature was averaged over 4 independent runs. These HNEMD calculations provide an additional bulk-transport validation complementary to the Green--Kubo results.

For interfacial transport, we carried out NEMD simulations in which heat was continuously injected into a hot slab and removed from a cold slab to establish a steady heat flux across the heterointerface. The simulations used a time step of 1.0~fs. At each target temperature \(T\), initial atomic velocities were assigned at \(T\), followed by 100~ps equilibration in the canonical ensemble. Steady-state heat transport was then generated using Langevin heat baths with a fixed temperature bias about the target temperature, such that \(T_{\mathrm{hot}}=T+\Delta T\) and \(T_{\mathrm{cold}}=T-\Delta T\), where \(\Delta T\) can be adjusted for different simulation conditions;  \(\Delta T=10\)~K is applied for investigating the length dependence. For the temperature-dependence test, \(\Delta T=20\)~K is applied because of the high-temperature conditions, and \(L=200\)~\AA{} is chosen considering the computational efficiency and the length-dependence. The hot and cold thermostat regions were each 10~\AA\ wide, and 10~\AA\ fixed slabs were imposed at both ends of the simulation cell. The remaining transport region was divided into bins of about 9--10~\AA\ width to compute the spatially averaged temperature profile. After a 500~ps warm-up under nonequilibrium conditions, production was continued for 4~ns, during which the temperature profile and heat flux were recorded. The thermal boundary conductance was obtained from the imposed heat flux divided by the temperature drop at the interface. These NEMD simulations were performed for bar lengths from 100 to 600~\AA\ (10 to 60~nm) and temperatures from 300 to 900~K, with each condition averaged over four independent runs.

To further analyze the microscopic origin of interfacial heat transfer, we performed spectral heat current (SHC) calculations on the relaxed interface models. In these calculations, atomic velocities and interatomic forces near the interface were sampled during the steady-state nonequilibrium simulations, and the frequency-resolved interfacial heat current was obtained from their time-correlation functions following the standard SHC formalism. In the representative setup used here, the system was first equilibrated for 200~ps in the canonical ensemble and then driven for 1~ns under nonequilibrium conditions to establish a steady temperature gradient before spectral analysis. The SHC calculation employed a sampling interval of 10~fs, a correlation length of 1000 samples, an output interval of 4000 steps, and 320 frequency points for the spectrum. The production trajectory used for SHC analysis was 4~ns long. In the input shown here, the SHC signal was evaluated for the selected interface-related group pair specified in the model definition; the exact group assignment depends on the structural labeling of the simulation cell. These SHC calculations were used to extract the frequency-resolved spectral conductance and related interfacial dynamical quantities discussed in the main text.

Structural models and atomistic simulation snapshots were visualized using OVITO \citep{Stukowski2010-cp}, while crystal structures, volumetric charge-density data, and related electronic-structure plots were visualized using VESTA \citep{Momma2011-bw}. All DFT/AIMD calculations, NEP training, molecular dynamics simulations, and post-processing analyses were carried out on a server equipped with 8 NVIDIA H100 GPUs. The typical GPU memory usage was approximately 10--20~GB per GPU. A typical 4-ns-long NEMD simulation with 25k atoms will take around 10--15 GPU hours.

Together, the EMD, HNEMD, NEMD, and SHC calculations test whether the simple feedforward neural network potential reproduces not only accurate fitting statistics, but also the bulk and interfacial transport physics needed for physically interpretable analysis of thermal conductance.

\section{Definitions of transport-analysis metrics}
\label{app:equations}

\subsection{Cross-interface coherence}
To analyze the microscopic origin of the conductance trends, we evaluate frequency-resolved spectral conductance, cross-interface coherence, and interfacial force-constant metrics. The cross-interface coherence is defined as the magnitude-squared coherence
\begin{equation}
C_{xy}(\omega)=\frac{|P_{xy}(\omega)|^2}{P_{xx}(\omega)P_{yy}(\omega)},
\end{equation}
where \(P_{xx}(\omega)\) and \(P_{yy}(\omega)\) are the auto-spectral densities of the selected vibrational signals on the SiC and Ga$_2$O$_3$ sides of the interface, respectively, and \(P_{xy}(\omega)\) is the corresponding cross-spectral density between the two sides. In the present calculation, these signals are taken as the near-interface atomic \(v_x\) fluctuations from the two slabs adjacent to the interface after subtracting the slab center-of-mass drift, and the reported coherence is obtained by averaging over matched left-right atomic pairs in the two interfacial regions. Thus, \(C_{xy}(\omega)\) measures the degree of frequency-resolved linear correlation, or phase-consistent vibrational coupling, across the interface. This definition follows the standard magnitude-squared coherence formalism and is used here as a dynamic descriptor of interfacial vibrational correlation, in the spirit of prior phonon-coherence analysis \cite{Carter1977-bs,Latour2014-lm}.

\subsection{Charge transfer and planar-averaged charge-density difference}

Interfacial charge transfer was quantified by Bader analysis\citep{Henkelman2006-nd} using the interface \texttt{POSCAR} and the atom-resolved Bader electron populations from \texttt{ACF.dat}. For each atom \(i\), the net charge was defined as \(q_i = Z_i^{\mathrm{val}} - N_i^{\mathrm{Bader}}\), where \(Z_i^{\mathrm{val}}\) is the valence electron count associated with the pseudopotential and \(N_i^{\mathrm{Bader}}\) is the Bader electron population. Atoms were assigned to the SiC or Ga$_2$O$_3$ side according to species, with passivating H excluded from the charge-transfer analysis. The total charge on each side was obtained by summing \(q_i\) over the corresponding atomic subset, and the interfacial charge-transfer magnitude was taken as \(|Q_{\mathrm{SiC}}|\) (equivalently \(|Q_{\mathrm{Ga_2O_3}}|\) within numerical precision). To compare different interface models, the charge transfer was also normalized by the interfacial area \(A\), evaluated from the cross product of the two in-plane lattice vectors.

To visualize interfacial charge redistribution, the planar-averaged charge-density-difference profile was computed from VASP \texttt{CHGCAR} files as
\[
\Delta \rho = \rho_{\mathrm{interface}} - \rho_{\mathrm{SiC}} - \rho_{\mathrm{Ga_2O_3}},
\]
with all charge densities evaluated on the same real-space grid. The charge-density data were processed using \textsc{pymatgen}, and \(\Delta\rho\) was averaged over planes parallel to the interface to obtain the one-dimensional redistribution profile along the interface-normal direction. The corresponding integrated profile was then obtained by cumulative integration of the planar-averaged \(\Delta\rho\). Together, the planar-averaged charge-density difference provides a spatial picture of electron accumulation and depletion near the interface, while the Bader analysis gives a scalar measure of the net charge transfer between the two sides.

\subsection{Harmonic force constants}
For lattice-dynamical analysis, harmonic force constants and phonon dispersions were computed using the finite-displacement method as implemented in \textsc{Phonopy} \citep{phonopy-phono3py-JPCM}. The second-order force constants were generated from NEP-evaluated forces on displaced supercells, and the resulting \texttt{FORCE\_CONSTANTS} matrices were used to calculate the phonon band structures along the high-symmetry paths reported in the main text. For SiC, a \(3\times3\times2\) supercell and a displacement amplitude of 0.02~\AA\ were used, whereas for \(\beta\)-Ga\(_2\)O\(_3\) a \(1\times3\times2\) supercell with the same 0.02~\AA\ displacement amplitude was adopted. This procedure provides a consistent lattice-dynamical validation of the trained potential in both constituent materials.

To quantify static interfacial vibrational coupling, cross-interface harmonic interatomic force-constant (IFC) metrics were extracted from the \textsc{Phonopy} force-constant matrices using a custom post-processing analysis. The cross-interface IFC blocks \(\Phi_{ij}\) were identified between atoms belonging to the SiC and Ga$_2$O$_3$ sides of the interface, where the interface position was determined from the species transition along the interface-normal direction. Interfacial atoms on both sides were selected within a 4 ~\AA-wide window on each side around the interface, and pairwise couplings were further filtered by a distance cutoff to exclude atoms too far from the interfacial region to contribute meaningfully to direct cross-interface coupling. For each retained atom pair \((i,j)\), the corresponding \(3\times 3\) harmonic IFC submatrix \(\Phi_{ij}\) was extracted, and the total cross-interface coupling strength was defined as
\begin{equation}
K_{\mathrm{cross}}=\sum_{i\in \mathrm{SiC}}\sum_{j\in \mathrm{Ga_2O_3}}\|\Phi_{ij}\|_F,
\end{equation}
where \(\|\Phi_{ij}\|_F\) denotes the Frobenius norm of the cross-interface IFC block. This quantity provides a compact scalar measure of the overall harmonic coupling strength across the interface, with larger values indicating that atomic displacements on one side generate stronger restoring forces on the other side.

To further resolve the directional character of the coupling, the IFC blocks were projected onto the interface normal and in-plane directions to obtain the normal and tangential coupling components, denoted \(K_{nn}\) and \(K_{tt}\), respectively. These projected quantities characterize how strongly the two sides of the interface are harmonically coupled through out-of-plane and in-plane vibrational motions. Because the two interface orientations considered in this work have different interfacial areas, all coupling metrics were additionally normalized by the interfacial area \(A\), yielding \(K_{\mathrm{cross}}/A\), \(K_{nn}/A\), and \(K_{tt}/A\). This normalization enables direct comparison between different interface structures on a common area basis. Together, these cross-interface IFC metrics serve as static descriptors of interfacial bonding and harmonic vibrational coupling, complementing the spectral conductance and coherence analyses by linking the observed transport behavior to the underlying interfacial force transmission pathways.

\subsection{Interfacial binding energy}
The interfacial binding energy was evaluated from the total energies of the relaxed interface supercell and the two corresponding isolated slab calculations using an area-normalized excess-energy definition. Specifically, the final interface energy \(E_{\mathrm{int}}\) was extracted from the interface calculation, while the slab reference energies \(E_{\mathrm{SiC}}\) and \(E_{\mathrm{Ga_2O_3}}\) were obtained from separate calculations for the isolated SiC and Ga$_2$O$_3$ slabs. To ensure a consistent comparison when the slab reference cells and interface cell contained different numbers of atoms, each slab energy was scaled according to the ratio of the number of non-excluded atoms of the corresponding material in the interface cell to that in the isolated slab cell, with passivating H atoms excluded from the scaling. The interfacial binding energy per unit area was then computed as\citep{Xu2023-interface}
\begin{equation}
\gamma_{\mathrm{int}}=\frac{E_{\mathrm{int}}-\left(s_{\mathrm{SiC}}E_{\mathrm{SiC}}+s_{\mathrm{Ga_2O_3}}E_{\mathrm{Ga_2O_3}}\right)}{A},
\end{equation}
where \(s_{\mathrm{SiC}}\) and \(s_{\mathrm{Ga_2O_3}}\) are the composition-based scaling factors for the SiC and Ga$_2$O$_3$ slab references, respectively, and \(A\) is the interfacial area. The resulting value was reported in both eV / \AA $^2$ and J/m$^2$ using the conversion factor \(1~\mathrm{eV/\AA^2}=16.02176565~\mathrm{J/m^2}\). More negative values of \(\gamma_{\mathrm{int}}\) indicate stronger interfacial adhesion and greater energetic stabilization upon interface formation.

\section{Additional temperature-dependent coupling analysis}
\label{app:coupling}
\begin{figure*}[t]
    \centering
    \includegraphics[width=0.95\textwidth]{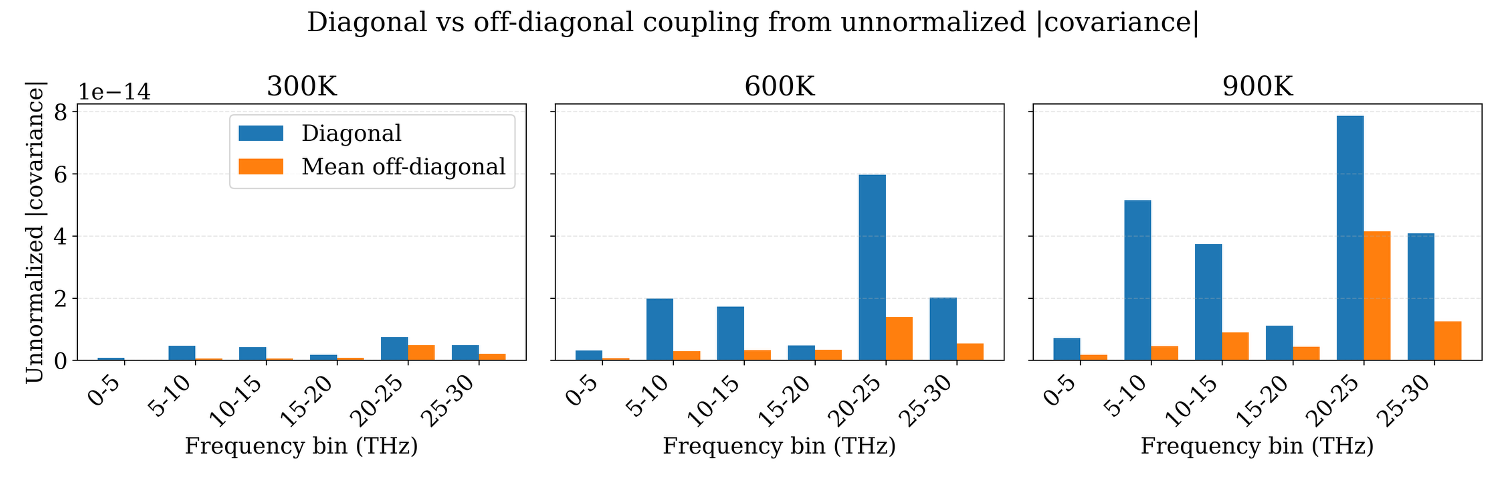}
    \caption{Supplementary bin-resolved cross-interface coupling analysis at \SIlist{300;600;900}{K}. The dominant channels remain broadly similar, but the relative strength of diagonal and off-diagonal interactions redistributes with temperature.}
    \label{fig8:appendix_coupling}
\end{figure*}

Figure~\ref{fig8:appendix_coupling} reports the binned cross-interface coupling derived from the unnormalized covariance of band-energy fluctuations between the two sides of the interface from 0 to 30 THz. Here, the diagonal terms measure same-bin coupling, whereas the off-diagonal terms represent the mean coupling to other frequency bins. 

The diagonal and off-diagonal terms represent two different ways that vibrational energy on one side of the interface can be dynamically linked to energy on the other side. In the band-resolved covariance analysis, a diagonal element corresponds to the coupling between the same frequency bin on the two sides of the interface, for example 5--10~THz on the SiC side correlated with 5--10~THz on the Ga$_2$O$_3$ side. Physically, this means that energy fluctuations remain within approximately the same spectral range during transmission. Such same-bin coupling is therefore most naturally associated with more frequency-conserving transfer and is qualitatively closer to coherent or elastic transport, although it is not a strict one-to-one measure of coherence.

By contrast, an off-diagonal element measures the coupling between different frequency bins across the interface, for example 5--10~THz on one side correlated with 10--15~THz or 20--25~THz on the other side. Its physical meaning is cross-bin energy exchange, namely that vibrational motion in one spectral range is dynamically linked to motion in another spectral range across the interface. This behavior is more naturally associated with frequency-mixing processes and therefore with incoherent, anharmonic, or inelastic exchange. In other words, the diagonal terms reflect how strongly similar-frequency vibrations communicate across the interface, whereas the off-diagonal terms reflect how strongly energy is redistributed between different frequency ranges during interfacial transfer. The covariance-based diagonal/off-diagonal decomposition is a complementary indicator: diagonal dominance suggests more same-bin, approximately elastic exchange, while increasing off-diagonal weight suggests stronger cross-bin mixing and thus a larger role of incoherent and anharmonic interfacial processes.

The dominant coupling channels remain broadly similar across temperature, indicating that the main transport window does not shift monotonically from one frequency region to another. What changes more clearly is the overall magnitude of the coupling, especially in the higher-frequency bins, where both diagonal and off-diagonal contributions increase with temperature. This pattern suggests stronger cross-bin vibrational mixing at elevated temperature and is therefore consistent with enhanced incoherent and anharmonic interfacial exchange. 

\section{Software assets and license compliance.} \label{app:assets}
\begin{table}[H]
\centering
\small
\caption{External software assets used in this work and their license or terms of use.}
\label{tab:software_licenses}
\begin{tabular}{p{0.13\linewidth} p{0.27\linewidth} p{0.50\linewidth}}
\toprule
Software & Role in this work & License / terms \\
\midrule
VASP & DFT/AIMD calculations & Proprietary; used under a valid institutional license in UW-Madison CMG group \\
LAMMPS & MD simulations & GPLv2 \\
GPUMD & NEP training / transport simulations & GPLv3 \\
Phonopy & Phonon and force-constant analysis & BSD 3-Clause \\
pymatgen & CHGCAR / structure processing & MIT \\
ASE & Structure conversion / preprocessing & LGPL \\
Bader code & Charge partitioning analysis & GPLv3 \\
OVITO & Atomistic visualization & MIT (OVITO Basic) \\
VESTA & Structure / volumetric visualization & Free for academic, scientific, educational, and noncommercial use \\
\bottomrule
\end{tabular}
\end{table}

\end{document}